\documentclass[letterpaper,11pt]{article}

\usepackage{amsmath,amssymb,array,caption,bbm}
\usepackage{graphicx,subfigure}
\usepackage{cite}
\usepackage[usenames,dvipsnames]{xcolor}
\usepackage{color}
\usepackage[unicode=true,pdfusetitle,
 bookmarks=true,bookmarksnumbered=false,bookmarksopen=false,citecolor=Turquoise,linktoc=page,
 breaklinks=false,pdfborder={0 0 1},backref=false,colorlinks=true,pdfpagemode=FullScreen]
 {hyperref}

\usepackage{colortbl}
\usepackage{tikz}
\usepackage{dcolumn}% Align table columns on decimal point
\usepackage{bm}% bold math 
\usepackage{multirow}

\usepackage{latexsym}
\usepackage{calc}
\usepackage{slashed}
\usepackage[mathscr]{eucal}
\usepackage{lineno}
\usepackage[normalem]{ulem}
\allowdisplaybreaks

\newcommand{\ignore}[1]{}
\renewcommand{\bf}{\textbf}

 \renewcommand{\t}{\mathrm{t}}

\newcommand{\h}{\hat}

\def\beq{\begin{equation}}
\def\eeq{\end{equation}}
\newcommand{\ba}{\begin{array}}
\newcommand{\ea}{\end{array}}
\newcommand{\bea}{\begin{eqnarray}}
\newcommand{\eea}{\end{eqnarray} }
\newcommand{\bal}{\begin{align}}
\newcommand{\eal}{\end{align}}
\def\bi{\begin{itemize}}
\def\ei{\end{itemize}}
\def\ben{\begin{enumerate}}
\def\een{\end{enumerate}}
\def\beq{\begin{equation}}
\def\eeq{\end{equation}}
\def\bc{\begin{center}}
\def\ec{\end{center}}
\def\bt{\begin{table}}
\def\et{\end{table}}
\def\btb{\begin{tabular}}
\def\etb{\end{tabular}}

\begin{document} 

\baselineskip=18pt

%%%%%%%%%%
%%%%%%%%%%    Title page
%%%%%%%%%%

\thispagestyle{empty}
\vspace{20pt}
\font\cmss=cmss10 \font\cmsss=cmss10 at 7pt

\begin{flushright}
\small 
\end{flushright}

\hfill
\vspace{20pt}

\begin{center}
{\Large \textbf
{Searching for Dark Photons with Maverick Top Partners
}}
\end{center}

\vspace{15pt}
\begin{center}
{\large Jeong Han Kim$\, ^{a}$, Samuel D. Lane$\, ^{a}$}, Hye-Sung Lee$\, ^{b}$, Ian M. Lewis$\, ^{a}$, Matthew Sullivan$\, ^{a}$
 \\
\vspace{15pt}
$^{a}$\textit{Department of Physics and Astronomy, University of Kansas, Lawrence, KS 66045, USA} \\
$^{b}$\textit{Department of Physics, KAIST, Daejeon 34141, Korea}

\end{center}

\vspace{20pt}

\begin{center}
\textbf{Abstract}
\end{center}
\vspace{5pt} {\small \noindent
In this paper, we present a model in which an up-type vector-like quark (VLQ) is charged under a new $U(1)_d$ gauge force which kinetically mixes with the SM hypercharge.  The gauge boson of the $U(1)_d$ is the dark photon, $\gamma_d$.  Traditional searches for VLQs rely on decays into Standard Model electroweak bosons $W,\,Z$ or Higgs.  However, since no evidence for VLQs has been found at the Large Hadron Collider (LHC), it is imperative to search for other novel signatures of VLQs beyond their traditional decays.    As we will show, if the dark photon is much less massive than the Standard Model electroweak sector, $M_{\gamma_d}\ll M_Z$, for the large majority of the allowed parameter space the VLQ predominately decays into the dark photon and the dark Higgs that breaks the $U(1)_d$ .  That is, this VLQ is a ``maverick top partner'' with nontraditional decays.  One of the appeals of this scenario is that  pair production of the VLQ at the LHC occurs through the strong force and the rate is determined by the gauge structure.  Hence, the production of the dark photon at the LHC only depends on the strong force and is largely independent of the small kinetic mixing with hypercharge.  This scenario provides a robust framework to search for a light dark sector via searches for heavy colored particles at the LHC. 
}
 
\vfill\eject
\noindent

%\tableofcontents
\newpage

%%%%%%%%%%%%%%%%%%%%%%%%%%%%%%%%%%%%%%%%%%%%%%%%%%%%%%
\section{Introduction}
%%%%%%%%%%%%%%%%%%%%%%%%%%%%%%%%%%%%%%%%%%%%%%%%%%%%%%
Some of the most important searches for Beyond the Standard Model (BSM) physics at the Large Hadron Collider (LHC) are searches for new vector like quarks (VLQs).  Up-type VLQs, so-called top partners $T$, are ubiquitous in composite~\cite{Agashe:2004rs,Agashe:2005dk,Agashe:2006at,Contino:2006qr,Giudice:2007fh,Azatov:2011qy,Serra:2015xfa} and Little Higgs~\cite{ArkaniHamed:2002qy,ArkaniHamed:2002pa,Low:2002ws,Chang:2003un,Csaki:2003si,Perelstein:2003wd,Chen:2003fm,Berger:2012ec} models where the top partners help solve the hierarchy problem.  Traditionally, searches for VLQs rely on decays into the Standard Model (SM) electroweak (EW) bosons: $W$/$Z$/Higgs.  However, there is a class of ``maverick top partners'' with non-traditional decays into photons~\cite{Kim:2018mks,Alhazmi:2018whk,DeRujula:1983ak,Kuhn:1984rj,Baur:1987ga,Baur:1989kv}, gluons~\cite{Kim:2018mks,Alhazmi:2018whk,DeRujula:1983ak,Kuhn:1984rj,Baur:1987ga,Baur:1989kv}, new scalars~\cite{Anandakrishnan:2015yfa,Dolan:2016eki,Bizot:2018tds,Chala:2018qdf, Aguilar-Saavedra:2017giu, Chala:2017xgc, Balkin:2017aep, Das:2019ivz}, etc. These new decays can easily be dominant with minor tweaks to the simplest VLQ models.  We consider a VLQ that is charged under both the SM and a new abelian gauge symmetry $U(1)_d$, where the SM is neutral under the $U(1)_d$.  As we will show, in a very large range of the parameter space, this opens new dominant decays of VLQs that have not yet been searched for. 

  This new $U(1)_d$ can be motivated by noting that dark matter may very well have self-interactions through this new force.  The $U(1)_d$ gauge boson, the so-called dark photon, kinetically mixes with the SM hypercharge through a renormalizable interaction~\cite{Holdom:1985ag,Bjorken:2009mm,Essig:2013lka}.  This kinetic mixing can be generated via new vector like fermions charged under both the SM and the new $U(1)_d$~\cite{Holdom:1985ag,Davoudiasl:2012ig,Lu:2017uur}, as considered here.  In the limit that the dark photon is much less massive than the Z and the kinetic mixing is small, the dark photon inherits couplings to SM particles of the form $\varepsilon \,J_{\rm EM}^\mu$, where $J_{\rm EM}^\mu$ is the electromagnetic current and $\varepsilon$ is the kinetic mixing parameter. Hence, the name dark photon.  Most of the searches for the dark photon take place at low energy experiments such as fixed target experiments or B-factories~\cite{Essig:2013lka}.  However, it is also possible to search for dark photons through the production and decays of heavy particles at high energy colliders~\cite{Curtin:2014cca,Kong:2014jwa,Alimena:2019zri}.  For example, Higgs decays~\cite{Davoudiasl:2012ig,Davoudiasl:2013aya,Lu:2017uur} into dark photons is a plausible scenario for discovery.

A recent paper~\cite{Rizzo:2018vlb} studied the scenario where down-type VLQs and vector like leptons are charged under the SM and $U(1)_d$.  Ref.~\cite{Rizzo:2018vlb} relied on a very large mass gap between the SM fermions and their vector-like fermion partners to suppress the traditional vector-like fermion decays into $W$/$Z$/Higgs.  With this mass gap, the branching ratios of vector-like fermions into dark photons and SM fermions is enhanced.  Here we point out that this mechanism does not require a mass gap in the fermion sector, although such a gap further enhances the effect.  To illustrate this, we will focus on an up-type VLQ, $T$, that mixes with the SM top quark, $t$. The mass gap between $t$ and $T$ does not need to be as large as between the bottom quark/leptons and vector-like fermions.  From the Goldstone equivalence theorem, the partial width of $T$ into fully SM final states is 
\begin{eqnarray}
\Gamma(T\rightarrow b/t+W/Z/h)\sim \sin^2\theta\,\frac{M_T^3}{v^2_{EW}}~,\nonumber
\end{eqnarray}
 where $\theta$ is a mixing angle between the SM top quark and $T$, $v_{EW}=246$~GeV is the Higgs boson vacuum expectation value (vev), and $M_T$ is the mass of the VLQ $T$.  The partial width is inversely proportional to $v^2_{EW}$ due to an enhancement of decays into longitudinal $W$s and $Z$s.  If the new vector like quark is charged under the dark force (and assuming there is a dark sector Higgs mechanism), the partial widths of $T$ into the dark photon, $\gamma_d$, or dark Higgs, $h_d$, is 
\begin{eqnarray}
\Gamma(T\rightarrow t+\gamma_d/h_d)\sim \sin^2\theta \frac{M_T^3}{v_d^2}~,\nonumber
\end{eqnarray}
where $v_d$ is the vev of the dark sector Higgs boson.  Note that now the partial width is inversely proportional to $v_d^2$.  Hence, the ratio of the rates into $\gamma_d/h_d$ and $W/Z/h$ is 
\begin{eqnarray}
\frac{\Gamma(T\rightarrow t+\gamma_d/h_d)}{\Gamma(T\rightarrow t/b+W/Z/h)}\sim \left(\frac{v_{EW}}{v_d}\right)^2~.\nonumber
\end{eqnarray}
  For dark photon masses $M_{\gamma_d}\lesssim 10$~GeV, we generically expect that the vev $v_d\lesssim 10$~GeV and 
\begin{eqnarray}
\frac{\Gamma(T\rightarrow t+\gamma_d/h_d)}{\Gamma(T\rightarrow t/b+W/Z/h)}\gtrsim \mathcal{O}(100)~.\nonumber
\end{eqnarray}
  Hence, the VLQ preferentially decays to light dark sector bosons due to the mass gap between the dark sector bosons and the SM EW bosons.  Since there is a quadratic dependence on $v_{EW}$ and $v_d$, this mass gap does not have to be very large for the decays $T\rightarrow t+\gamma_d/h_d$ to be dominant.

This is a new avenue to search for light dark sectors using decays of heavy particles at the LHC, providing a connection between heavy particle searches and searches for new light sectors.  The appeal of such searches is that pair production of VLQs is through the QCD interaction and is fully determined via $SU(3)$ gauge interactions. That is, the pair production rate only depends on the mass, spin, and color representation of the produced particles.  Additionally, as we will show, for a very large region of parameter space VLQs will predominantly decay into dark photons and dark Higgses.  Hence, the dark photon can be produced at QCD rates at the LHC independently of a small kinetic mixing parameter.  The major dependence on the kinetic mixing parameter $\varepsilon$ arises in the decay length of the dark photon, and for small $\varepsilon$ the dark photon can be quite long lived. In fact, for small dark photon masses, its decay products will be highly collimated and may give rise to displaced ``lepton jets''~\cite{ArkaniHamed:2008qp}. 

In Section~\ref{sec:model} we present an explicit model that realizes this mechanism for dark photon searches and review current constraints in Sec.~\ref{sec:Const}.   We calculate the decay and production rates of the new VLQ in Sec.~\ref{sec:proddec} and the decay of the dark photon in Sec.~\ref{sec:Adecay}. In Section~\ref{sec:collider} we present collider searches relevant for our model.   This includes the current collider sensitivity as well as demonstrating the complementarity between the searches for dark photons via heavy particle decays at the LHC and low energy experiments.  Finally we conclude in Section~\ref{sec:conc}.  We also include several appendices with the details on perturbative unitarity calculation in App.~\ref{app:PertUnit}, kinetic mixing in App.~\ref{app:KineticM}, relevant scalar interactions in App.~\ref{app:Scalar}, and relevant dark photon-fermion couplings in App.~\ref{app:Dark}.

%%%%%%%%%%%%%%%%%%%%%%%%%%%%%%%%%%%%%%%%%%%%%%%%%%%%%%
\section{Model}
\label{sec:model}
%%%%%%%%%%%%%%%%%%%%%%%%%%%%%%%%%%%%%%%%%%%%%%%%%%%%%%

We consider a simple extension of the SM consisting of a new $SU(2)_L$ singlet up-type vector-like quark, $t_{2}$, and a new $U(1)_d$ gauge symmetry. For simplicity, we will only consider mixing between the new vector-like quark and $3^{\text{rd}}$ generation SM quarks:
\begin{eqnarray}
Q_{L}=\begin{pmatrix} t_{1L} \\ b_L \end{pmatrix}~,\quad t_{1R}~,\quad {\rm and}\quad b_R~.
\end{eqnarray}
The SM particles are singlets under the new symmetry, and we give the VLQ $t_2$ a charge $+1$ under the new symmetry. The $U(1)_d$ is broken by a dark Higgs field $H_d$ that is a singlet under the SM and has charge $+1$ under $U(1)_d$.  The relevant field content and their charges under $SU(3) \times SU(2)_L \times U(1)_Y \times U(1)_d$ are given in Table \ref{tab:particles}. This particle content and charges are similar to those in Ref. \cite{Rizzo:2018vlb}.

This field content allows for kinetic mixing between the SM $U(1)_Y$ field, $B'_{\mu}$, and the new $U(1)_d$ gauge boson, $B'_{d,\mu}$:
\begin{eqnarray} \label{eq:GaugeK}
\mathcal{L}_{\text{Gauge}}&=&-\frac{1}{4}G^A_{\mu\nu}G^{A,\mu\nu}-\frac{1}{4}W^a_{\mu\nu}W^{a,\mu\nu}-\frac{1}{4}B'_{\mu\nu}{B'}^{\mu\nu}\\ \nonumber
&+&\frac{\varepsilon'}{2\cos\hat{\theta}_W} B'_{d,\mu\nu}{B'}^{\mu\nu} - \frac{1}{4}B'_{d,\mu\nu}{B'_d}^{\mu\nu}~,
\end{eqnarray}
where $G^A_{\mu \nu}$ are the $SU(3)$ field strength tensor with $A = 1, ... , 8$ and $W^a_{\mu \nu}$ are the $SU(2)_L$ field strength tensors with $a = 1, 2, 3.$ The relevant fermion kinetic terms for the third generation quarks and VLQ are
\begin{eqnarray}
\mathcal{L}_{F,\,kin}=\overline{Q}_Li\slashed{D}Q_L+\overline{t}_{1R}i\slashed{D}t_{1R}+\overline{b}_Ri\slashed{D}b_R+\overline{t}_2i\slashed{D}t_2~,
\end{eqnarray}
and the relevant scalar kinetic terms are
\begin{eqnarray} \label{eq:ScalarK}
\mathcal{L}_{S,kin}=|D_\mu\Phi|^2+|D_\mu H_d|^2~.
\end{eqnarray}
The general covariant derivative is
\begin{eqnarray} \label{eq:Cov}
D_\mu=\partial_\mu-ig_St^AG_\mu^A-ig T^aW^a_\mu-ig' Y B'_\mu-i g'_d Y_d B'_{d,\mu}~,
\end{eqnarray}
where $g_S$ is the strong coupling constant, $g$ is the $SU(2)_L$ coupling constant, $g'$ is the $U(1)_Y$ coupling constant, and $g'_d$ is the $U(1)_d$ coupling constant. Values for $Y$, $Y_d$ and the generators of $SU(3)$, $SU(2)_L$ are given according to the charges in Table \ref{tab:particles}.

\begin{table}[t]
\begin{center}
\begin{tabular}{|c|c|c|c|c|}\hline
 &\; $SU(3)$ \;&\; $SU(2)_L$ \;&\; $Y$ \;&\; $Y_d \;$\\\hline\hline
$t_{1R}$ & {\bf 3} & {\bf 1} & 2/3 & 0\\\hline
$b_R$ & {\bf 3} & {\bf 1} & -1/3 & 0\\\hline
\; $Q_{L}=\begin{pmatrix} t_{1L} \\ b_L \end{pmatrix}$ \;& {\bf 3} & {\bf 2} & 1/6 & 0\\\hline
$\Phi$ & {\bf 1} & {\bf 2} & 1/2 & 0\\\hline
$t_{2L}$ & {\bf 3} & {\bf 1} & 2/3 & 1\\\hline
$t_{2R}$ & {\bf 3} & {\bf 1} & 2/3 & 1\\\hline
$H_d$ & {\bf 1} & {\bf 1} & 0 & 1\\\hline\hline
\end{tabular}
\end{center}
\caption{\label{tab:particles} Field content and their charges. $t_{1R}$, $b_R$, and $Q_L$ are $3^{\text{rd}}$ generation SM quarks, $\Phi$ is the SM Higgs doublet, $t_{2}$ is the $SU(2)_L$ singlet VLQ, and $H_d$ is the $U(1)_d$ Higgs field. $Y$ is the SM Hypercharge and $Y_d$ is the $U(1)_d$ charge. }
\end{table}

%%%%%%%%%%%%%%%%%%%%%%%%%%%%%%%%%%%%%%%%%%%%%%%%%%%%%%%%%%%%%%%%
\subsection{Scalar Sector}
%%%%%%%%%%%%%%%%%%%%%%%%%%%%%%%%%%%%%%%%%%%%%%%%%%%%%%%%%%%%%%%%
%
The allowed form of the scalar potential symmetric under the gauge group $SU(3)_C \times SU(2)_L \times U(1)_Y \times U(1)_d$ is
\begin{eqnarray} \label{eq:HiggsPo}
V(\Phi,H_d)=-\mu^2|\Phi|^2+\lambda\,|\Phi|^4-\mu^2_{h_d}|H_d|^2+\lambda_{h_d}|H_d|^4+\lambda_{hh_d}|\Phi|^2|H_d|^2~.
\end{eqnarray}
Since $H_d$ does not break EW symmetry, $\Phi$ must have a vev of $v_{EW}=246$~GeV.  Imposing that the potential has a minimum where the SM Higgs and dark Higgs have vacuum expectation values $\langle \Phi\rangle = (0,v_{EW}/\sqrt{2})^t$ and $\langle H_d\rangle = v_d/\sqrt{2}$, the mass parameters are found to be
\begin{eqnarray}
\mu^2=\lambda\,v^2_{EW}+\frac{\lambda_{hh_d}}{2}v_d^2~,\quad 
\mu_{h_d}^2=\lambda_{h_d}\,v_d^2+\frac{\lambda_{hh_d}}{2}v_{EW}^2~.
\end{eqnarray}

Now we work in the unitary gauge:
\begin{eqnarray}
\Phi=\frac{1}{\sqrt{2}}\begin{pmatrix} 0\\v_{EW}+h\end{pmatrix},\quad H_d=\frac{1}{\sqrt{2}}(v_d+h_d)~.
\end{eqnarray}
The two Higgs bosons $h,\,h_d$ mix and can be rotated to the mass basis:
\begin{eqnarray}
\begin{pmatrix}h_1\\h_2\end{pmatrix}=\begin{pmatrix}\cos\theta_S&-\sin\theta_S\\ \sin\theta_S&\cos\theta_S \end{pmatrix}\begin{pmatrix}h\\h_d\end{pmatrix}~,
\end{eqnarray}
where $h_1$ can be identified as the observed Higgs boson with a mass $M_1 = 125$~GeV, and $h_2$ is a new scalar boson with mass $M_2$. After diagonalizing the mass matrix, the free parameters of the scalar sector are
\begin{eqnarray}
\theta_S,\quad M_1=125~{\rm GeV},\quad M_2,\quad v_d, \quad{\rm and}\quad v_{EW}=246~{\rm GeV}~.
\end{eqnarray}
All parameters in the Lagrangian can then be determined
%\begin{eqnarray}
%
\begin{gather}
\displaystyle\lambda=\frac{M_1^2\cos^2\theta_S+M_2^2\sin^2\theta_S}{2\,v^2_{EW}}~,\quad \lambda_{h_d}=\frac{M_2^2\cos^2\theta_S+M_1^2\sin^2\theta_S}{2\,v^2_{d}}~,\nonumber\\
\displaystyle\lambda_{hh_d}=\frac{M_2^2-M_1^2}{2\,v_{EW}\,v_d}\sin2\theta_S~,\nonumber\\
\displaystyle \mu^2=\frac{1}{2}\left[M_1^2\cos^2\theta_S+M_2^2\sin^2\theta_S+\frac{\tan\beta}{2}\left(M_2^2-M_1^2\right)\sin2\theta_S\right]~,\nonumber\\
\displaystyle\mu_{h_d}^2=\frac{1}{2}\left[M_2^2\cos^2\theta_S+M_1^2\sin^2\theta_S+\frac{1}{2\,\tan\beta}\left(M_2^2-M_1^2\right)\sin2\theta_S\right]~,\label{eq:PotParam}
%\end{array}
\end{gather}
where $\tan \beta = v_d /v_{EW}$.

To check the stability of the scalar potential, we consider it when the fields $\Phi$ and $H_d$ are large:
\begin{eqnarray}
V(\Phi,H_d)&\rightarrow& \lambda\,|\Phi|^4+\lambda_{hh_d}|\Phi|^2|H_d|^2+\lambda_{h_d}|H_d|^4\nonumber\\
&=&\left(\lambda-\frac{1}{4}\frac{\lambda_{hh_d}^2}{\lambda_{h_d}}\right)|\Phi|^4+\lambda_{h_d}\left(|H_d|^2+\frac{1}{2}\frac{\lambda_{hh_d}}{\lambda_{h_d}}|\Phi|^2\right)^2~,
\end{eqnarray}
where in the last step we completed the square.  The potential is bounded when
\begin{eqnarray}
4\,\lambda_{h_d}\lambda\geq \lambda_{hh_d}^2,\quad \lambda>0,\quad {\rm and}\quad \lambda_{h_d}>0~.\label{eq:Bounded}
\end{eqnarray}
From the relationships in Eq.~(\ref{eq:PotParam}) we have
\begin{eqnarray}
4\,\lambda_{h_d}\lambda-\lambda_{hh_d}^2=\frac{M_1^2M_2^2}{v_d^2v_{EW}^2}>0~.
\end{eqnarray}
Hence, the boundedness condition for the potential is always satisfied as long as $\lambda$ and $\lambda_{h_d}$ are both positive.

For our analysis in the next sections, only two trilinear scalar couplings are relevant:
\begin{eqnarray}
V(h_1,h_2)\supset \frac{1}{2}\lambda_{122}\,h_1\,h_2^2+\frac{1}{2}\lambda_{112}\,h_1^2\,h_2~,
\end{eqnarray}
where
\begin{eqnarray}
\lambda_{122}&=&-\frac{M_1^2+2\,M_2^2}{2\,v_d}\sin 2\theta_S\left(\cos\theta_S-\tan\beta\,\sin\theta_S\right)~,\\
\lambda_{112}&=&\frac{2\,M_1^2+M_2^2}{2\,v_d}\sin 2\theta_S\left(\tan\beta\,\cos\theta_S+\sin\theta_S\right)~.\nonumber
\end{eqnarray}

%%%%%%%%%%%%%%%%%%%%%%%%%%%%%%%%%%%%%%%%%%%%%%%%%%%%%%%%%%%%%%%%
\subsection{Gauge Sector}
%%%%%%%%%%%%%%%%%%%%%%%%%%%%%%%%%%%%%%%%%%%%%%%%%%%%%%%%%%%%%%%%
From Eq.(\ref{eq:GaugeK}), the $U(1)_d$ gauge boson can mix with the SM electroweak gauge bosons. After diagonalizing the gauge bosons, the covariant derivative in Eq.~(\ref{eq:Cov}) becomes
\begin{eqnarray}
D_\mu &=&\partial_\mu-ig_S t^A G^A_\mu-igT^+W^+-ig T^-W^--ieQA_\mu\label{eq:CovComplete}\\
&&-i\left(\hat{g}_Z\hat{Q}_Z\cos\theta_d - g_d Y_d \sin\theta_d -\varepsilon \frac{g'}{\cos\hat{\theta}_W}Y\sin\theta_d\right)Z_\mu\nonumber\\
&&-i\left(\hat{g}_Z\hat{Q}_Z\sin\theta_d + g_d Y_d \cos\theta_d + \varepsilon \frac{g'}{\cos\hat{\theta}_W} Y \cos\theta_d\right) \gamma_{d,\mu}~ ,\nonumber
\end{eqnarray}
where $\theta_d$ is a mixing angle between the dark photon and SM $Z$-boson; $e=g\sin\hat\theta_W=g'\cos\hat\theta_W$ and $Q = T^3 + Y$ are the usual electromagnetic charge and operator respectively; $\hat{g}_Z=e/\cos\hat\theta_W/\sin\hat\theta_W$ and $\hat{Q}_Z=T_3-\hat{x}_W Q$ with $\hat{x}_W=\sin^2\hat{\theta}_W$ are the neutral current coupling and operator respectively; $T^{\pm} = (T^1 \pm i T^2) / \sqrt{2}$; $Z$ is the observed EW neutral current boson with mass $M_Z$; and $\gamma_{d, \mu}$ is the dark photon with mass $M_{\gamma_d}$.  Additionally, $\hat{\theta}_W$ is the mixing angle between $B'_\mu$ and $W^3_\mu$.  The relationship between $\hat{\theta}_W$ and other model parameters is not the same as the SM weak mixing angle.  Hence, we introduce the hat notation to emphasize the difference.  The relationship between the SM weak mixing angle and $\hat{\theta}_W$ is given below\footnote{The details of diagonalizing the gauge boson kinetic and mass terms, including precise definitions of $\hat{\theta}_W$ and $\theta_d$, are given in Appendix \ref{app:KineticM}.}. For simplicity of notation, we have redefined the coupling constant and kinetic mixing parameter
\begin{eqnarray}
g_d = g'_d / \sqrt{1 - \varepsilon'^2 /\cos^2 \hat{\theta}_W}\quad{\rm and}\quad \varepsilon = \varepsilon' / \sqrt{ 1 - \varepsilon'^2 / \cos^2 \hat{\theta}_W}~.
\end{eqnarray}

The SM EW sector has three independent parameters, which we choose to be the experimentally measured $Z$ mass, the fine-structure constant at zero momentum, and G-Fermi \cite{Tanabashi:2018oca}:
\begin{eqnarray}
M_Z = 91.1876~\text{GeV},~ \alpha^{-1}_{EM} (0) = 137.035999074 , ~G_F = 1.1663787 \times 10^{-5}~\text{GeV}^{-2}~ .\label{eq:EWParms}
\end{eqnarray} 
In addition to the EW parameters, we have the new free parameters:
\begin{eqnarray}
v_d,\quad M_{\gamma_d},\quad{\rm and},\quad \varepsilon~.
\end{eqnarray}
All other parameters in the gauge sector can be expressed in terms of these. Since $\varepsilon^2\ll 1$, we can solve equations for $\sin\theta_d,\,g_d,$ and $\cos\hat{\theta}_W$ iteratively as an expansion in $\varepsilon$:
\begin{eqnarray} \nonumber
\sin\theta_d&=&\frac{\tan\theta^{SM}_W}{1-\tau_{\gamma_d}^2}\varepsilon+\mathcal{O}(\varepsilon^3),\quad\cos\hat{\theta}_W=\cos\theta_W^{SM}+\mathcal{O}(\varepsilon^2),\quad{\rm and}\quad
g_d=\frac{M_{\gamma_d}}{v_d}+\mathcal{O}(\varepsilon^2)~,
\end{eqnarray}
where $\tau_{\gamma_d}=M_{\gamma_d}/M_Z$ and the SM value of the weak mixing angle is
\begin{eqnarray}
\cos^2{\theta}_W^{SM}=\frac{1}{2}+\frac{1}{2}\sqrt{1-\frac{2\sqrt{2}\,\pi \alpha_{EM}(0) }{G_F\,M_Z^2}} ~.
\end{eqnarray}
Although these are good approximations for $\varepsilon^2\ll 1$, unless otherwise noted we will use exact expressions of parameters as given in Appendix~\ref{app:KineticM}.

Note that in the limit of small kinetic mixing $\varepsilon\ll 1$ and dark photon mass much less than the $Z$-mass $M_{\gamma_d} \ll M_Z$ we find the covariant derivative

\begin{eqnarray} \label{eq:NewCovar}
D_\mu &=&\partial_\mu-ig_S t^A G^A_\mu-ig^{SM}T^+W^+-ig^{SM} T^-W^--ieQA_\mu\\
&&-i\left[g^{SM}_ZQ^{SM}_Z-\varepsilon g_d\,Y_d\,\tan\theta^{SM}_W\right]Z_\mu-i\left[\varepsilon e Q+g_d Y_d\right]\gamma_{d,\mu}+\mathcal{O}(\varepsilon^2,M_{\gamma_d}^2/M_Z^2) ~, \nonumber
\end{eqnarray}
where the superscript $SM$ indicates the SM value of parameters.  Hence we see that the dark photon couples to SM particles through the electromagnetic current with coupling strength $\varepsilon e Q$.  Additionally, the $Z$-boson obtains additional couplings to particles with non-zero dark charge $Y_d$ with strength $\varepsilon\,g_d\,Y_d\,\tan\theta^{SM}_W$.

%%%%%%%%%%%%%%%%%%%%%%%%%%%%%%%%%%%%%%%%%%%%%%%%%%%%%%%%%%%%%%%%
\subsection{Fermion Sector}
%%%%%%%%%%%%%%%%%%%%%%%%%%%%%%%%%%%%%%%%%%%%%%%%%%%%%%%%%%%%%%%%
%
To avoid flavor constraints, we only allow the VLQ $t_2$ to mix with the third generation SM quarks. The allowed Yukawa interactions and mass terms that are symmetric under $SU(3) \times SU(2)_L \times U(1)_Y \times U(1)_d$ are
\begin{eqnarray}\label{eq:FermionLag}
\mathcal{L}_{Yuk}= -y_b \overline{Q}_L \Phi b_{R} -y_t \overline{Q}_L\widetilde{\Phi}t_{1R} - \lambda_t H_d \overline{t}_{2L} t_{1R} - M_{t_2} \overline{t}_{2L} t_{2R}+{\rm h.c.} ~.
\end{eqnarray}
After symmetry breaking, the VLQ and top quark mass terms are then 
\begin{eqnarray}
\mathcal{L}_{T,mass}=-\overline{\chi}_L \mathcal{M} \chi_R+{\rm h.c.}~,
\end{eqnarray}
where
\begin{eqnarray}
\displaystyle\chi_\tau =\begin{pmatrix} t_{1\tau}\\t_{2\tau}\end{pmatrix},~\mathcal{M}=\begin{pmatrix} \displaystyle\frac{y_t\,v_{EW}}{\sqrt{2}} & 0 \\ \displaystyle\frac{\lambda_t\,v_d}{\sqrt{2}} & \displaystyle M_{t_2} \end{pmatrix}~,\label{eq:topmass}
\end{eqnarray}
and $\tau=L,\,R$.  To diagonalize the mass matrix, we perform the bi-unitary transformation:
\begin{eqnarray}
\begin{pmatrix}t_L \\ T_L\end{pmatrix} = \begin{pmatrix} \cos\theta^t_L & -\sin\theta^t_L \\ \sin\theta^t_L & \cos\theta^t_L \end{pmatrix}\begin{pmatrix} t_{1L}\\ t_{2L}\end{pmatrix}~,\quad\begin{pmatrix}t_R \\ T_R\end{pmatrix} = \begin{pmatrix} \cos\theta^t_R & -\sin\theta^t_R \\ \sin\theta^t_R & \cos\theta^t_R \end{pmatrix}\begin{pmatrix} t_{1R}\\ t_{2R}\end{pmatrix}~,
\end{eqnarray}
where $t$ and $T$ are the mass eigenstates with masses $M_t=173$~GeV and $M_T$, respectively.  Since the Lagrangian only has three free parameters, the top sector only has three inputs which we choose to be
\begin{eqnarray}
M_t=173~{\rm GeV},\quad M_T,\quad{\rm and}\quad \theta^t_L~.
\end{eqnarray}
The Lagrangian parameters $\lambda_t,\,y_t,\,M_{t_2}$ can be expressed by
\begin{eqnarray}
y_t&=&\sqrt{2}\frac{\sqrt{M_t^2\cos^2\theta^t_L+M_T^2\sin^2\theta^t_L}}{v_{EW}}~,\\
\lambda_t&=&\frac{(M_T^2-M_t^2)\sin2\theta^t_L}{\sqrt{2}v_d\sqrt{M_t^2\cos^2\theta^t_L+M_T^2\,\sin^2\theta^t_L}}~,\label{eqn:lambda_t}\\
M_{t_2}&=&\frac{M_t\,M_T}{\sqrt{M_t^2\cos^2\theta^t_L+M_T^2\sin^2\theta^t_L}}~.\label{eq:Mt2}
\end{eqnarray}
The right-handed mixing angle is redundant and can be determined via
\begin{eqnarray}
\cos\theta^t_R=\frac{M_{t_2}}{M_T}\cos\theta^t_L\quad{\rm and}\quad \sin\theta^t_R=\frac{M_{t_2}}{M_t}\sin\theta^t_L~.\label{eq:thetaR}
\end{eqnarray}

 After rotating to the scalar and fermion mass eigenbases, the $h_{1,2}$ couplings to the third generation and VLQ are given by
\begin{eqnarray} \label{eq:Top_Scalar}
\mathcal{L} &\supset&- h_1 \left[\lambda_{tt}^{h_1} \overline{t}t+\lambda_{TT}^{h_1}\overline{T}T+\overline{t}\left(\lambda_{tT}^{h_1} P_R+\lambda_{Tt}^{h_1} P_L\right)T+\overline{T}\left(\lambda_{Tt}^{h_1} P_R+\lambda_{tT}^{h_1} P_L\right)t\right]\\
&-& h_2 \left[\lambda_{tt}^{h_2} \overline{t}t+\lambda_{TT}^{h_2}\overline{T}T+\overline{t}\left(\lambda_{tT}^{h_2}P_R+\lambda_{Tt}^{h_2} P_L\right)T+\overline{T}\left(\lambda_{Tt}^{h_2} P_R+\lambda_{tT}^{h_2} P_L\right)t\right] ~,\nonumber
\end{eqnarray}
where the $h_1$ couplings are
\bea
\lambda_{tt}^{h_1} &=& \displaystyle\frac{1}{\sqrt{2}}\cos \theta^t_R \left(y_t\cos \theta^t_L \cos \theta_S + \lambda_t \sin \theta^t_L \sin \theta_S \right)~ ,\label{eq:h1Yuk}\\
 \lambda_{tT}^{h_1}&=&\displaystyle\frac{1}{\sqrt{2}}\sin \theta^t_R \left(y_t \cos \theta^t_L \cos \theta_S + \lambda_t \sin \theta^t_L \sin \theta_S \right)~,\label{eq:htt} \nonumber \\
\lambda_{Tt}^{h_1}&=&\displaystyle\frac{1}{\sqrt{2}}\cos \theta^t_R \left(y_t \sin \theta^t_L \cos \theta_S - \lambda_t \cos \theta^t_L \sin \theta_S\right) ~, \nonumber \\
\lambda_{TT}^{h_1}&=&\displaystyle\frac{1}{\sqrt{2}}\sin \theta^t_R \left(y_t \sin \theta^t_L \cos \theta_S - \lambda_t \cos \theta^t_L \sin \theta_S \right)~,\nonumber
\eea
and the $h_2$ couplings are
\bea
\lambda_{tt}^{h_2} &=& \displaystyle\frac{1}{\sqrt{2}}\cos \theta^t_R \left(y_t\cos \theta^t_L \sin \theta_S - \lambda_t \sin \theta^t_L \cos \theta_S \right) ~,\label{eq:h2Yuk}\\
 \lambda_{tT}^{h_2}&=&\displaystyle\frac{1}{\sqrt{2}}\sin \theta^t_R \left(y_t \cos \theta^t_L \sin \theta_S - \lambda_t \sin \theta^t_L \cos \theta_S \right)~,\label{eq:h2tt} \nonumber \\
\lambda_{Tt}^{h_2}&=&\displaystyle\frac{1}{\sqrt{2}}\cos \theta^t_R \left(y_t \sin \theta^t_L \sin \theta_S + \lambda_t \cos \theta^t_L \cos \theta_S\right) ~, \nonumber \\
\lambda_{TT}^{h_2}&=&\displaystyle\frac{1}{\sqrt{2}}\sin \theta^t_R \left(y_t \sin \theta^t_L \sin \theta_S + \lambda_t \cos \theta^t_L \cos \theta_S \right)~.\nonumber
\eea

Now we consider the small angle limit $|\theta_L^t|\ll1$. If the VLQ and top quark have similar masses $M_T\sim M_t$, then Eq.~(\ref{eq:Mt2}) becomes $M_{t_2}\sim M_T\sim M_t$.  In this limit, from Eq.~(\ref{eq:thetaR}), we see that $\theta_L^t\sim \theta_R^t$ and both mixing angles are small.  However, for a large fermion mass hierarchy $M_t/M_T\ll1$, the right-handed mixing angle expressions in Eq.~(\ref{eq:thetaR}) become
\begin{eqnarray}
\cos\theta^t_R\approx\frac{M_t/M_T}{\sqrt{\sin^2\theta_L^t+M_t^2/M_T^2}},\quad\sin\theta^t_R\approx\frac{\sin\theta_L^t}{\sqrt{\sin^2\theta_L^t+M_t^2/M_T^2}}~.\label{eq:thetaRlim}
\end{eqnarray}
There are two cases then:
\begin{eqnarray}
\sin\theta_R^t\sim \begin{cases} (M_T/M_t)\sin\theta_L^t& \text{if}~|\sin\theta_L^t|<M_t/M_T\ll 1\\
\pm 1 & \text{if}~M_t/M_T\lesssim |\sin\theta_L^t|\ll1~,\end{cases}
\end{eqnarray}
where the sign of $\pm1$ depends on the sign of $\theta_L^t$.  Hence, as discussed in Ref.~\cite{Rizzo:2018vlb}, the right-handed mixing angle is enhanced relative to the left-handed mixing angle due to a large fermion mass hierarchy.

Since $t_{2}$ and $t_{1}$ have different quantum numbers and mix, flavor off-diagonal couplings between the VLQ $T$ and the SM third generation quarks appear.  In the small mixing angle limit, $M_t/M_T,|\theta_L^t|,|\varepsilon|\ll1$ the relevant couplings are
\begin{eqnarray}
\begin{array}{ll}
W-T-b & \displaystyle \sim i\frac{g}{\sqrt{2}}\sin\theta_L^t\,\gamma^\mu\,P_L~,\\[2ex]
Z-T-t & \displaystyle \sim i\frac{g_Z^{SM}}{2}\sin\theta_L^t\,\gamma^\mu\,P_L+ig_d\,\frac{(M_T/M_t)\sin\theta_L^t}{1+(M_T/M_t)^2\sin^2\theta_L^t}\sin\theta_d\,\gamma^\mu\,P_R~,\\[2ex]
\gamma_d-T-t&\displaystyle \sim -i\,g_d\,\sin\theta_L^t\,P_L-i\,g_d\,\frac{(M_T/M_t)\sin\theta_L^t}{1+(M_T/M_t)^2\sin^2\theta_L^t}P_R~.
\end{array}\label{eq:Gauge_Top}
\end{eqnarray}
The full Feynman rules are given in Appendix~\ref{app:Dark}.  
Note that although the right-handed coupling to the $Z$ appears of order $\theta^2$, if $M_t/M_T\sim |\theta_L^t|\sim |\theta_d|$ the left- and right-handed couplings can be of the same order.  However, with this counting the right-handed coupling of the dark photon, VLQ, and top quark is unsuppressed.  This is precisely the fermionic mass hierarchy enhancement noticed in Ref.~\cite{Rizzo:2018vlb}.  However, as we will point out, a fermionic mass hierarchy is not necessary for the VLQ decays into the dark Higgs or dark photon to be dominant.

%%%%%%%%%%%%%%%%%%%%%%%%%%%%%%%%%%%%%%%%%%%%%%%%%%%%%%
\section{Current Constraints}
\label{sec:Const}
%%%%%%%%%%%%%%%%%%%%%%%%%%%%%%%%%%%%%%%%%%%%%%%%%%%%%%

%%%%%%%%%%%%%%%%%%%%%%%%%%%%%%%%%%%%%%%%%%%%%%%%%%%%%%%%%%%%%%%%
\subsection{Electroweak Precision and Direct Searches}
%%%%%%%%%%%%%%%%%%%%%%%%%%%%%%%%%%%%%%%%%%%%%%%%%%%%%%%%%%%%%%%%
Electroweak precision measurements place strong constraints on the addition of new particles. In the model presented here, there are many contributions to the oblique parameters~\cite{Peskin:1991sw,Degrassi:1992ue,Degrassi:1993kn}: new loop contributions from the VLQ~\cite{Lavoura:1992np,He:2001tp,Aguilar-Saavedra:2013qpa,Dawson:2012di,Ellis:2014dza,Chen:2017hak} and scalar~\cite{Bowen:2007ia,Profumo:2007wc,Barger:2007im,Pruna:2013bma,Lopez-Val:2014jva,Robens:2015gla} as well as shifts in couplings to EW gauge boson couplings from the mixing of the dark photon with hypercharge~\cite{Hook:2010tw,Curtin:2014cca}, dark Higgs with the SM Higgs, and the VLQ with the top quark.  Since there are multiple contributions to the oblique parameters in this model, there is the possibility of cancellations that could relax some of the constraints.  To be conservative, we will only consider one contribution at a time.

\begin{figure}[tb]
\begin{center}
\subfigure[]{\includegraphics[width=0.47\textwidth,clip]{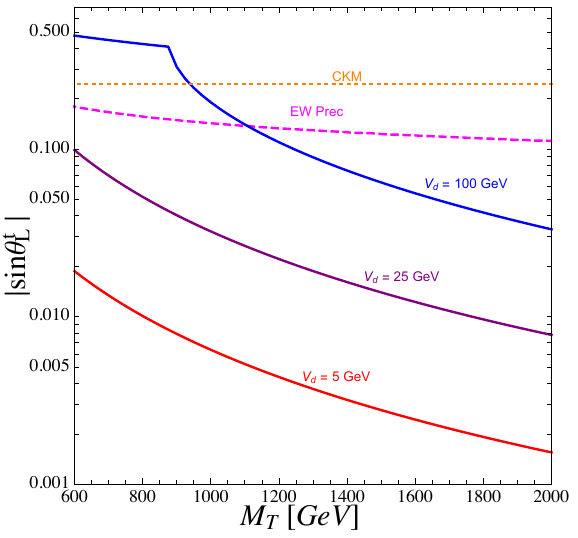} \label{thetaL-MT-plane}}
\subfigure[]{\includegraphics[width=0.45\textwidth,clip]{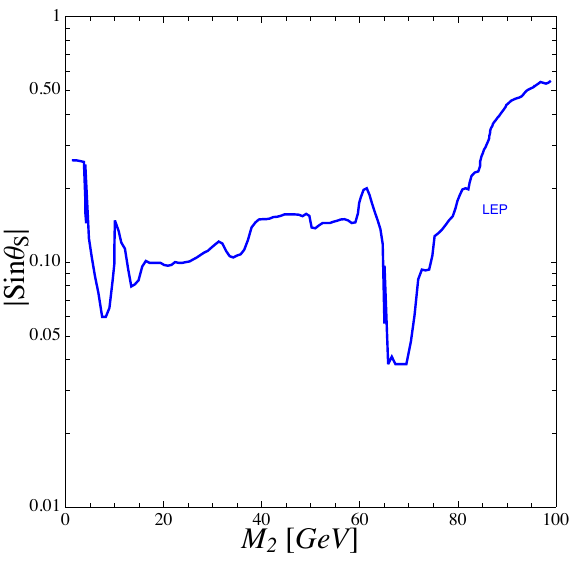} \label{thetas-M2-plane}}
\caption{  (a) Upper bounds for $|\sin\theta_L^t|$ from (dashed magenta) EW precision data from Ref.~\cite{Chen:2017hak}; (dotted orange) current CKM measurements; and (solid) requiring $\lambda_t$ satisfy Eq.~(\ref{eq:lamtbnd}) for (red) $v_d=5$~GeV, (violet) $v_d=25$~GeV, and (blue) $v_d=100$~GeV.  (b) Upper bound  on $|\sin\theta_S|$ from LEP for $M_2<100$~GeV as found in Ref.~\cite{Robens:2015gla}. 
}
\end{center}
\end{figure}

There are also many direct searches for VLQs, new scalars, and dark photons at colliders and fixed target experiments.  Here we summarize the current state of constraints:

\begin{itemize}
\item {\bf {VLQ:}} The dashed line labeled ``EW Prec'' in Fig.~\ref{thetaL-MT-plane} shows the EW precision constraints on the VLQ-top quark mixing angle. This result is taken from Ref.~\cite{Chen:2017hak}.  The current limits are $|\sin\theta_L^t|\lesssim 0.16$ ($0.11$) for VLQ mass $M_T=1$ TeV (2 TeV).

Additionally, in our model the top-bottom component of the CKM matrix is $V_{tb} = (V_{tb})_{SM} \cos \theta_L^t$, where the subscript SM denotes the SM value. The most stringent constraints on $V_{tb}$ come from single top quark production.  A combination of Tevatron and LHC single top measurements give a constraint of $|V_{tb}|=1.019 \pm 0.025$~\cite{Tanabashi:2018oca}. Another more recent analysis including differential distributions gives a bound of $|V_{tb}|=0.986 \pm 0.008$~\cite{Clerbaux:2018vup}. Both constraints give an upper bound of  $|\sin \theta_L^t| \le 0.24$ at the $95 \% $ confidence level. This limit is indicated by the orange dotted line labeled ``CKM'' in Fig.~\ref{thetaL-MT-plane}, where the region above is excluded. We see that the CKM measurements are not currently as important as EW precision constraints.

As mentioned above, in the model presented here traditional $T$ decays into SM EW bosons $Z,\,W,$ Higgs will be suppressed and not directly applicable. Nevertheless, for completeness we summarize their results here.  In these traditional modes, the LHC excludes VLQ masses $M_T\lesssim 1.1-1.4$~TeV in pair production searches~\cite{Aaboud:2018pii,CMS-PAS-B2G-18-005,Sirunyan:2018omb} and $M_T\lesssim 1-1.2$~TeV in single production searches~\cite{Aaboud:2018saj,Sirunyan:2017ynj,Sirunyan:2017tfc}.  Single production of an $SU(2)_L$ singlet $T$ depends on the mixing angle $\theta_L^t$ and decouples as $\theta_L^t\rightarrow 0$~\cite{Kim:2018mks} weakening the above limit.  Taking this into account, LHC searches for single $T$ production have been cast into constraints on $\theta_L^t$ which are comparable to EW precision constraints for $M_T\lesssim 1$~TeV~\cite{Aaboud:2018ifs,Aaboud:2018saj}.  

\item {\bf {Scalar:}} The addition of a new scalar shifts Higgs boson couplings away from SM predictions, as well as contributing to new loop contributions to EW precision parameters.  Additionally, many searches have been performed for new scalar production at the LHC~\cite{Khachatryan:2015cwa, Sirunyan:2018qlb, Aaboud:2017rel, Aad:2014yja, Khachatryan:2016sey, Bechtle:2015pma, Haisch:2018kqx} as well as at LEP~\cite{Bechtle:2008jh, Bechtle:2011sb, Bechtle:2013wla}.  However, the most stringent constraints~\cite{test1} come from precision measurements of the observed $M_1=125$~GeV Higgs boson for $M_1\lesssim M_2\lesssim 650$~GeV and precision $W$-mass constraints~\cite{Lopez-Val:2014jva,Robens:2015gla,Ilnicka:2018def} for $650~{\rm GeV}\lesssim M_2\lesssim 1$~TeV.  The constraints on the scalar mixing angle is $|\sin\theta_S|\lesssim 0.21-0.22$ for $M_1<M_2< 1$~TeV~\cite{test1}.  For $M_2<100$~GeV LEP searches can be very constraining on the scalar mixing angle, as shown in Fig.~\ref{thetas-M2-plane}.   These results are adapted from Ref.~\cite{Robens:2015gla}. 

\item {\bf {Kinetic Mixing:}} As can be seen in covariant derivative in Eq.~(\ref{eq:CovComplete}), the couplings between the $Z$ and SM particles are shifted due to the kinetic mixing of the Hypercharge and $U(1)_d$ gauge boson.  Hence, electroweak precision data can place bounds on the value of the kinetic mixing parameter $\varepsilon$~\cite{Hook:2010tw,Curtin:2014cca}.  The most stringent constraints from EW precision are $|\varepsilon|\lesssim 3\times10^{-2}$~\cite{Curtin:2014cca}.  This is less constraining than direct searches for dark photons at fixed target experiments or low energy experiments~\cite{Battaglieri:2017aum} which require $|\varepsilon|\lesssim 10^{-3}$ for $M_{\gamma_d}=0.1-10$~GeV.
\end{itemize} 

%%%%%%%%%%%%%%%%%%%%%%%%%%%%%%%%%%%%%%%%%%%%%%%%%%%%%%%%%%%%%%%%
\subsection{Perturbativity Bounds}
%%%%%%%%%%%%%%%%%%%%%%%%%%%%%%%%%%%%%%%%%%%%%%%%%%%%%%%%%%%%%%%%

Requiring the top quark and VLQ Yukawa couplings be perturbative can place strong constraints on the top quark-VLQ mixing angle.  As can be seen in Eq.~(\ref{eqn:lambda_t}), in the limit that $M_t/M_T\sim|\sin\,\theta_L^t|\ll1$ the Yukawa couplings become
\begin{eqnarray}
y_t&=&\sqrt{2}\frac{M_t}{v_{EW}}\sqrt{1+\frac{M_T^2}{M_t^2}\sin^2\theta_L^t}+\mathcal{O}(\sin^2\theta_L^t,M_t^2/M_T^2),\nonumber\\
\lambda_t&=&\sqrt{2}\frac{M^2_T-M_t^2}{v_d M_t}\sin\theta_L^t+\mathcal{O}(\sin^2\theta_L^t)~.
\end{eqnarray}
While $y_t$ is well-behaved for $M_t/M_T\sim|\sin\theta_L^t|$, $\lambda_t$ is enhanced by $M_T/v_d$.  Hence, the mixing angle must be small to compensate for this and ensure $\lambda_t$ remains perturbative.

To determine when $\lambda_t$ becomes non-perturbative, we calculate the perturbative unitarity limit for the $H_dt\rightarrow H_dt$ scattering process and find that 
\begin{eqnarray}
|\lambda_t|\leq 4\sqrt{2 \pi}~.\label{eq:pertunit}
\end{eqnarray}
When this limit is saturated, there must be a minimum higher order correction of $41\%$ to unitarize the S-matrix~\cite{Schuessler:2007av}.  Hence, this is near or at the limit for which we can trust perturbative calculations.  Details of this calculation can be found in Appendix~\ref{app:PertUnit}.

To translate the limit on $\lambda_t$ to a limit on the mixing angle $\sin\theta_L^t$ we solve Eq.~(\ref{eqn:lambda_t}) to find
\begin{eqnarray}
|\sin\theta_L^t|=\frac{1}{2}\sqrt{\frac{2\,M_T^2-2\,M_t^2-\lambda_t^2v_d^2}{M_T^2-M_t^2}\left(1-\sqrt{1-\frac{8\,\lambda_t^2 v_d^2 M_t^2}{(2\,M_T^2-2\,M_t^2-v_d^2\lambda_t^2)^2}}\right)}~.
\end{eqnarray}
This solution is real if $|\lambda_t|\leq \sqrt{2}(M_T-M_t)/v_d$.  Combining with the perturbative unitarity limit in Eq.~(\ref{eq:pertunit}), we find an upper limit on $\lambda_t$:
\begin{eqnarray}
|\lambda_t| \leq \sqrt{2}\min\left\{\frac{M_T-M_t}{v_d},4\sqrt{\pi} \right\}~.\label{eq:lamtbnd}
\end{eqnarray}
Note that for VLQ mass $M_T< 4\sqrt{\pi}\,v_d+M_t$, the perturbative unitarity limit is never saturated.  Hence, for a fixed $v_d$ there is an upper bound on $M_T$ for which $\lambda_t$ is always perturbative.  Assuming $M_t,\,v_d\ll M_T$, the upper-bound on $\sin\theta_L^t$ becomes
\begin{eqnarray}
|\sin\theta_L^t|\lesssim\begin{cases} \displaystyle 4\sqrt{\pi}\frac{v_d\,M_t}{M_T^2}&\text{for}~M_T\geq 4\sqrt{\pi}\,v_d+M_t\\
\sqrt{M_t/M_T}&\text{for}~M_T< 4\sqrt{\pi}\,v_d+M_t\end{cases}~.\label{eq:sthLtBND}
\end{eqnarray}

In Fig.~\ref{thetaL-MT-plane} we show the limits on $|\sin\theta_L^t|$ from (solid) requiring that $\lambda_t$ satisfies Eq.~(\ref{eq:lamtbnd}) for various values of $v_d$ together with (dashed magenta) EW precision data and (dotted orange) CKM constraints.  The kink in the $v_d=100$~GeV line occurs at VLQ mass $M_T\sim 4\sqrt{\pi}\,v_d+M_t\sim 880$~GeV.  For $M_T< 4\sqrt{\pi}\,v_d+M_t$ the upper bound on $\sin\theta_L^t$ is proportional to $M_T^{-1/2}$, while for $M_T\geq 4\sqrt{\pi}v_d+M_t$ it is proportional to $M_T^{-2}$ as shown in Eq.~(\ref{eq:sthLtBND}).  As can be clearly seen, over much of the parameter range the limits on $\lambda_t$ in Eq.~(\ref{eq:lamtbnd}) provide the most stringent constraint on $\sin\theta_L^t$.  As mentioned earlier, this is due to $\lambda_t$ having an enhancement of $M_T^2/M_t/v_d$, requiring $\sin\theta_L^t$ to be quite small to ensure $\lambda_t$ does not get too large.  EW precision is more constraining for larger $v_d$ and smaller $M_T$.

%%%%%%%%%%%%%%%%%%%%%%%%%%%%%%%%%%%%%%%%%%%%%%%%%%%%%%%%%%%%%%%%
\subsection{$h_1\rightarrow \gamma_d \gamma_d$ Limits}
%%%%%%%%%%%%%%%%%%%%%%%%%%%%%%%%%%%%%%%%%%%%%%%%%%%%%%%%%%%%%%%%

There have been searches at the LHC~\cite{Aaboud:2018fvk} for $h_1\rightarrow \gamma_d\gamma_d\rightarrow 4\ell$ where $\ell=e,\mu$ that place limits on combination
\begin{eqnarray}
\frac{\sigma(pp\rightarrow h_1)}{\sigma_{SM}(pp\rightarrow h_1)}{\rm BR}(h_1\rightarrow \gamma_d \gamma_d)\lesssim BR_{\lim}~,\label{eq:BRlimOrig}
\end{eqnarray}
for dark photons in the mass range $1~{\rm GeV}<M_{\gamma_d}<60~{\rm GeV}$. The subscript $SM$ indicates a SM production rate.  The $h_1$ production rate is dominantly via gluon fusion which in the model presented here is altered via the shift in the $h_1-t-t$ coupling away from the SM prediction as shown in Eqs.~(\ref{eq:Top_Scalar},\ref{eq:h1Yuk}) and new loop contributions from the new VLQ.  However, in the small mixing angle limit with the counting $\theta_L^t\sim \theta_S\sim M_t/M_T$, we have 
\begin{eqnarray}
\sigma(pp\rightarrow h_1)=\sigma_{SM}(pp\rightarrow h_1)+\mathcal{O}(\theta^2)~.
\end{eqnarray}

In addition to the usual SM decay modes, $h_1$ can decay into $\gamma_d\gamma_d$, $\gamma_d Z$, and $h_2h_2$ when kinematically allowed.  Using the counting $\varepsilon\sim\theta_L^t\sim \theta_S\sim M_{\gamma_d}/M_Z\sim M_2/M_Z$, the partial widths into the new decay modes are
\begin{eqnarray}
\Gamma(h_1\rightarrow h_2h_2)&\approx&\Gamma(h_1\rightarrow \gamma_d\gamma_d)=\frac{M_1^3\sin^2\theta_S}{32\,\pi\,v_d^2}+\mathcal{O}(\theta^3)~,\label{eq:h1togdgd}\\
\Gamma(h_1\rightarrow Z\gamma_d)&=&\mathcal{O}(\theta^4)~.\nonumber
\end{eqnarray}

For the decays into SM, all the couplings between $h_1$ and SM fermions and gauge bosons, except for the $h_1-Z-Z$ and $h_1-t-t$ couplings, are uniformly suppressed by $\cos\theta$.  The $h_1-Z-Z$ and $h_1-t-t$ couplings are more complicated due to the $Z-\gamma_d$ mixing and $t-T$ mixing, respectively.  Additionally, there are new contributions to the loop level decays $h_1\rightarrow gg$, $h_1\rightarrow \gamma\gamma$, and $h_1\rightarrow \gamma Z$ due to the new VLQ.  Since the partial widths $\Gamma(h_1\rightarrow \gamma\gamma)$ and $\Gamma(h_1\rightarrow Z\gamma)$ make negligible contributions to the total width, we will neglect changes in these quantities.  Reweighting the SM partial widths with the new contributions, the width into fully SM final states are then
\begin{eqnarray}
\Gamma(h_1\rightarrow X_{SM}X^{(*)}_{SM})&=&\cos^2\theta_S\,\left(\Gamma_{SM}(h_1\rightarrow X_{SM} X_{SM}^{(*)})-\Gamma_{SM}(h_1\rightarrow ZZ^*)-\Gamma_{SM}(h_1\rightarrow gg)\right)\nonumber\\
&&+\left(\cos\theta_S-\frac{g_d^2\,v_{EW}\,v_d}{M_Z^2}\sin^2\theta_d\left(\cos\theta_S+\sin\theta_S\frac{v_{EW}}{v_d}\right)\right)^2\Gamma_{SM}(h_1\rightarrow ZZ^*)\nonumber\\
&&+\left|\frac{v_{EW}\,\lambda_{tt}^{h_1}}{M_t}-\frac{4}{3}\frac{v_{EW}\,\lambda_{TT}^{h_1}}{M_T\,F(\tau_t)}\right|^2\Gamma_{SM}(h_1\rightarrow gg)\\
&=&\Gamma_{SM}(h_1)+\mathcal{O}(\theta^2)~,
\end{eqnarray}
where $X_{SM}$ are SM fermions or gauge bosons, the subscript $SM$ indicates SM values of widths, $\Gamma_{SM}(h_1)=4.088$~MeV~\cite{deFlorian:2016spz}, and $\lambda_{tt}^{h_1}$, $\lambda_{TT}^{h_1}$ are in Eq.~(\ref{eq:h1Yuk}).  Other SM values for the partial widths of $h_1$ can be found in Ref.~\cite{deFlorian:2016spz}.  The loop function $F(\tau_i)$ can be found in Ref.~\cite{Gunion:1989we}, where $\tau_i=4\,M_i^2/M_1^2$ and we have used $M_T\gg M_1$ such that $F(\tau_T)=F(\infty)=-4/3$.  

The total width is then
\begin{eqnarray}
\Gamma_{tot}(h_1)&=&\Gamma(h_1\rightarrow X_{SM}X^{(*)}_{SM})+\Gamma(h_1\rightarrow \gamma_d\gamma_d)+\Gamma(h_1\rightarrow Z \gamma_d)+\Gamma(h_1\rightarrow h_2h_2)\theta(M_1-2\,M_2)\nonumber\\
&=&\Gamma_{SM}(h_1)+\mathcal{O}(\theta^2)~,
\end{eqnarray}
and Eq.~(\ref{eq:BRlimOrig}) becomes
\begin{eqnarray}
BR_{lim}\geq \frac{\sigma(pp\rightarrow h_1)}{\sigma_{SM}(pp\rightarrow h_1)}{\rm BR}(h_1\rightarrow \gamma_d \gamma_d)=\frac{\Gamma(h_1\rightarrow \gamma_d\gamma_d)}{\Gamma_{SM}(h_1)}+\mathcal{O}(\theta^4)~.
\end{eqnarray}
Using Eq.~(\ref{eq:h1togdgd}) we find the limit 
\begin{eqnarray}
|\sin\theta_S|\leq \sqrt{\frac{32\,\pi\,v_d^2\Gamma_{SM}(h_1)}{M_1^3}BR_{lim}}=4.6\times 10^{-4}\left(\frac{v_d}{\rm GeV}\right)\sqrt{BR_{lim}}~.\label{eq:sthlim}
\end{eqnarray}

\begin{figure}[!htb!]
\begin{center}
\subfigure[]{\includegraphics[width=0.49\textwidth,clip]{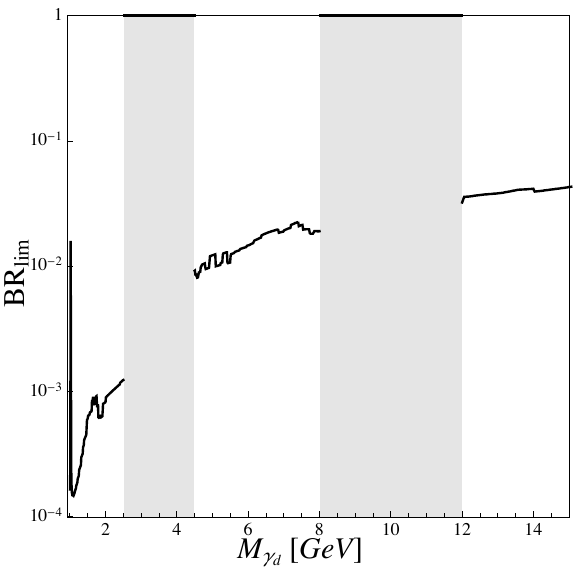}\label{fig:BRlim}}
\subfigure[]{\includegraphics[width=0.49\textwidth,clip]{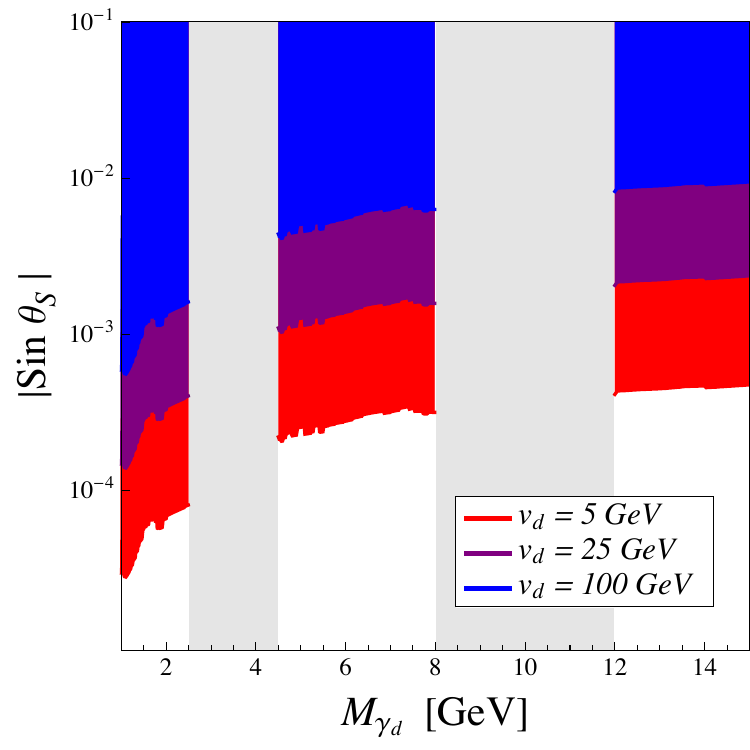}\label{fig:sthlim}}
\caption{(a) $BR_{lim}$ as defined in Eq.~(\ref{eq:BRlimOrig}).  These are the ATLAS results for a $h_1\rightarrow 2\gamma_d\rightarrow 4\mu$ search~\cite{Aaboud:2018fvk} reweighted by ${\rm BR}(\gamma_d\rightarrow \mu^+\mu^-)$ including hadronic decays. The dashed regions are not included in the $h_1\rightarrow 2\gamma_d\rightarrow 4\mu$ search due to resonant hadrons~\cite{Aaboud:2018fvk}. (b) Upper bounds on $|\sin\theta_S|$ from Eq.~(\ref{eq:sthlim}).  The solid colored regions are ruled out for (red) $v_d=5$~GeV, (maroon) $v_d=25$ GeV, and (blue) $v_d=100$~GeV.}
\end{center}
\end{figure}

ATLAS has measured the upper limit $BR_{lim}$ in the mass range $M_{\gamma_d}=1-15$~GeV when both dark photons decay into muons~\cite{Aaboud:2018fvk}.  However, they have assumed ${\rm BR}(\gamma_d\rightarrow e^-e^+)={\rm BR}(\gamma_d\rightarrow \mu^-\mu^+)=0.5$ neglecting possible hadronic decays of the dark photon.  We reweight the results of Ref.~\cite{Aaboud:2018fvk} using the ${\rm BR}(\gamma_d\rightarrow \mu^+\mu^-)$ including hadronic decays\footnote{See Sec.~\ref{sec:Adecay} for details of the ${\rm BR}(\gamma_d\rightarrow \mu^+\mu^-)$ calculation.}, as shown in Fig.~\ref{fig:BRlim}.   The hatched regions correspond to hadronic resonances and were not included in the search in Ref.~\cite{Aaboud:2018fvk}. This is the $BR_{lim}$ to be used in Eqs.~(\ref{eq:BRlimOrig},\ref{eq:sthlim}).

In Fig.~\ref{fig:sthlim} we show the upper limit on $\sin\theta_S$ from Eq.~(\ref{eq:sthlim}) and using $BR_{lim}$ in Fig.~\ref{fig:BRlim}.  The solid regions are ruled out by the $h_1\rightarrow 2\gamma_d\rightarrow 4\mu$ search for (red) $v_d=5$~GeV, (maroon) $v_d=25$~GeV, and (blue) $v_d=100$~GeV.  These constraints are very strong with limits in the range of $|\sin\theta_S|\lesssim 10^{-5}-10^{-2}$.  These limits are more constraining than the direct searches for $h_2$ as shown in Fig.~\ref{thetas-M2-plane}.  Eq.~(\ref{eq:sthlim}) is linear in the dark Higgs vev $v_d$, so the limits on $\sin\theta_S$ become less constraining for large $v_d$.  However, since $M_{\gamma_d}\approx g_d\,v_d$ these constraints cannot be arbitrarily relaxed without very small dark gauge coupling $g_d$.

If there is dark matter (DM) with mass $M_{DM}< M_{\gamma_d}/2$, it is possible that the decay of the dark photon into DM is dominant since, unlike the dark photon coupling to SM fermions, the $\gamma_d$-DM coupling would not be suppressed by the kinetic mixing parameter $\varepsilon$.  Hence, it is possible for the Higgs to decay invisibly $h_1\rightarrow 2\gamma_d\rightarrow {\rm DM}$.  There are searches for invisible decays of $h_1$ with limits ${\rm BR}(h_1\rightarrow {\rm Invisible})\leq 0.19$~\cite{Sirunyan:2018owy} from CMS and ${\rm BR}(h_1\rightarrow {\rm Invisible})\leq0.26$ from ATLAS~\cite{ATLAS-CONF-2018-054}.   Assuming that ${\rm BR}(\gamma_d\rightarrow {\rm DM})=1$, from Eq.~(\ref{eq:sthlim}) these limits correspond to
\begin{eqnarray}
|\sin\theta_S|\leq \left(\frac{v_d}{\rm GeV}\right)\times\begin{cases}
		2.0\times 10^{-4} &\text{for~CMS~\cite{Sirunyan:2018owy}}\\
		2.3\times 10^{-4} &\text{for~ATLAS~\cite{ATLAS-CONF-2018-054}}~.\end{cases}
\end{eqnarray}

%%%%%%%%%%%%%%%%%%%%%%%%%%%%%%%%%%%%%%%%%%%%%%%%%%%%%%
\section{Production and Decay of Vector Like Quark}
\label{sec:proddec}
%%%%%%%%%%%%%%%%%%%%%%%%%%%%%%%%%%%%%%%%%%%%%%%%%%%%%%
%
\begin{figure}[!htb!]
\begin{center}
\subfigure[]{\includegraphics[width=0.28\textwidth,clip]{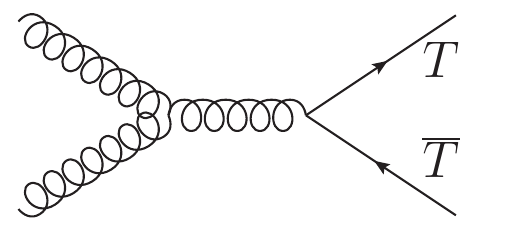}\label{fig:ggTTs}}
\subfigure[]{\includegraphics[width=0.20\textwidth,clip]{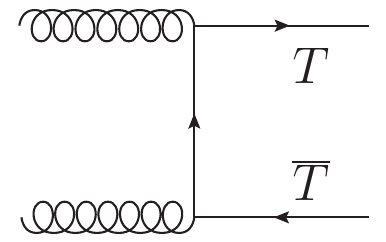}\label{fig:ggTTt}}
\subfigure[]{\includegraphics[width=0.28\textwidth,clip]{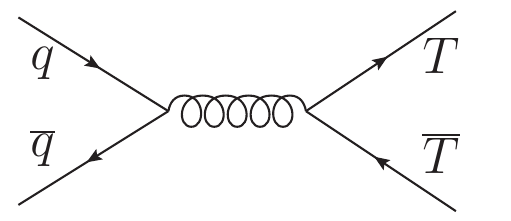}\label{fig:qqTTs}}\\
\subfigure[]{\includegraphics[width=0.23\textwidth,clip]{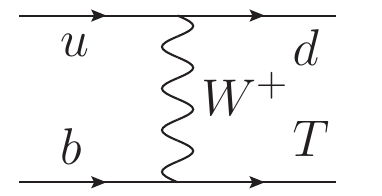}\label{fig:Tjet}}
\subfigure[]{\includegraphics[width=0.23\textwidth,clip]{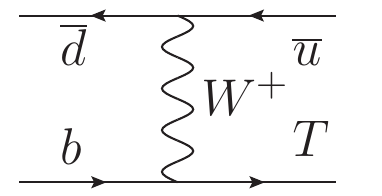}\label{fig:Tjet1}}
\end{center}
\caption{Standard production modes of VLQs at the LHC for (a-c) pair production and (d,e) VLQ plus a jet production.  The conjugate processes for (d,e) are not shown here.\label{fig:Prod_T}}
\end{figure}

\begin{figure}[t]
\begin{center}
\includegraphics[width=0.49\textwidth,clip]{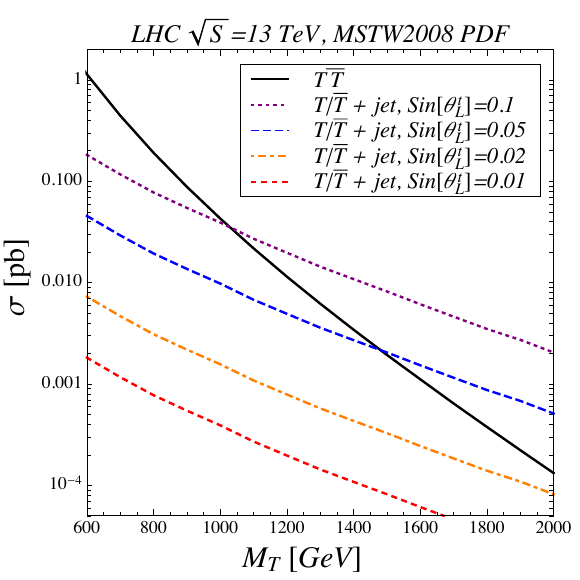}
\caption{\label{fig:T_prod_13}  Pair $T\bar{T}$ and single $T+$jet production cross sections~\cite{Matsedonskyi:2014mna}. Pair production is at NNLO in QCD and single production at NLO in QCD.
}
\end{center}
\end{figure}

In this section, we focus on the production and decay of the VLQ, $T$, at the LHC based on the model in Sec.~\ref{sec:model}. Figure \ref{fig:Prod_T} displays the VLQ (a,b,c) pair production ($T \overline{T}$) and (d,e) single production in association with a jet ($T/\overline{T}$ + jet)\footnote{There is also $T W^{-} + \overline{T} W^{+}$ production which is subdominant. In the model with an additional $SU(2)_L$ singlet scalar, a loop-induced $T \overline{t} + \overline{T} t$ production \cite{Kim:2018mks} can be as large as the pair production.}. The pair production is induced by QCD interactions so that the production cross section depends only on $M_T$, the spin of $T$, and the gauge coupling.  Hence, pair production is relatively model independent\footnote{In a scenario where the top partners are pair produced via a heavier resonance, the production cross section can be model dependent. See Refs. \cite{Chala:2014mma, Azatov:2015xqa, Araque:2015cna} and references therein.}. The single production, on the other hand, relies on the $b-W-T$ coupling in Eq.(\ref{eq:Gauge_Top}) which is proportional to the mixing angle $\sin \theta^t_L$. Therefore the production cross section is proportional to $\sin^2 \theta^t_L$ and is suppressed for small $\theta_L^t$~\cite{Kim:2018mks}.

In Fig.~\ref{fig:T_prod_13} we show cross sections for single and pair production of $T$ from Ref.~\cite{Matsedonskyi:2014mna}\footnote{It should be noted that these results are for a charge 5/3 VLQ. However, a charge 2/3 partner has the same QCD and spin structure so the results are still valid since the QCD production does not depend on the electric charge of the particle.}. The pair production cross section with NNLO QCD corrections is computed using the HATHOR code \cite{Aliev:2010zk} with the MSTW2008 parton distribution functions (PDF) \cite{Martin:2009iq}. The single production cross section with NLO QCD corrections is calculated using MCFM \cite{Campbell:2009gj, Campbell:2009ss,  Campbell:2004ch} with the same PDF. The NLO single production cross sections are rescaled by $\sin^2 \theta^t_L$ to take into account the normalization of the $b-W-T$ coupling in Eqs.(\ref{eq:Gauge_Top},\ref{eq:charged}). The single production becomes more important at high mass, where the gluon PDF sharply drops suppressing $gg\rightarrow T\overline{T}$ and the pair production phase space is squeezed relative to single production.  With a sizable mixing angle $|\sin \theta^t_L| \gtrsim 0.1$, the single production outperforms the pair production in a wide range of $M_T$.  The single production, however, vanishes as the $t-T$ mixing angle becomes very small, as required by perturbativity and EW precision [Fig.~\ref{thetaL-MT-plane}]. This can be already seen from Figure \ref{fig:T_prod_13} when $\sin \theta^t_L = 0.01$, where the $T+$jet cross section goes into the sub-femtobarn level which will be challenging to probe at the LHC.

\begin{figure}[!htb!]
\begin{center}
\subfigure[]{\includegraphics[width=0.28\textwidth,clip]{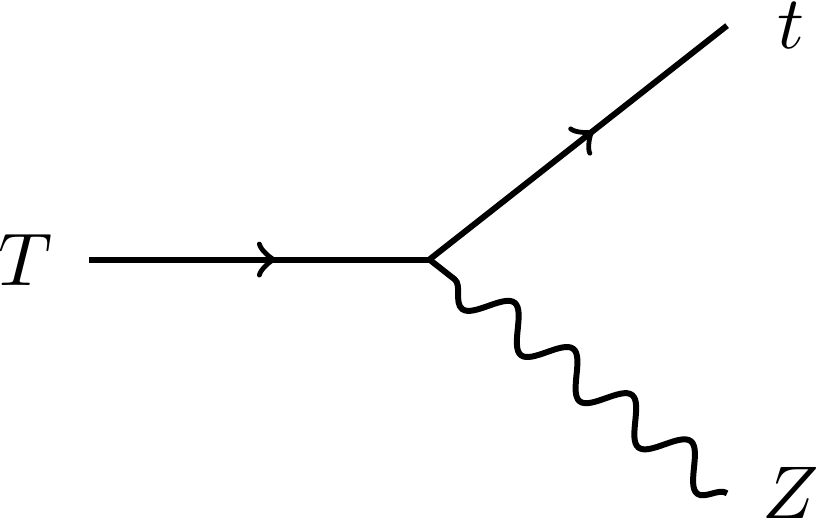}\label{fig:TtZ}}
\subfigure[]{\includegraphics[width=0.28\textwidth,clip]{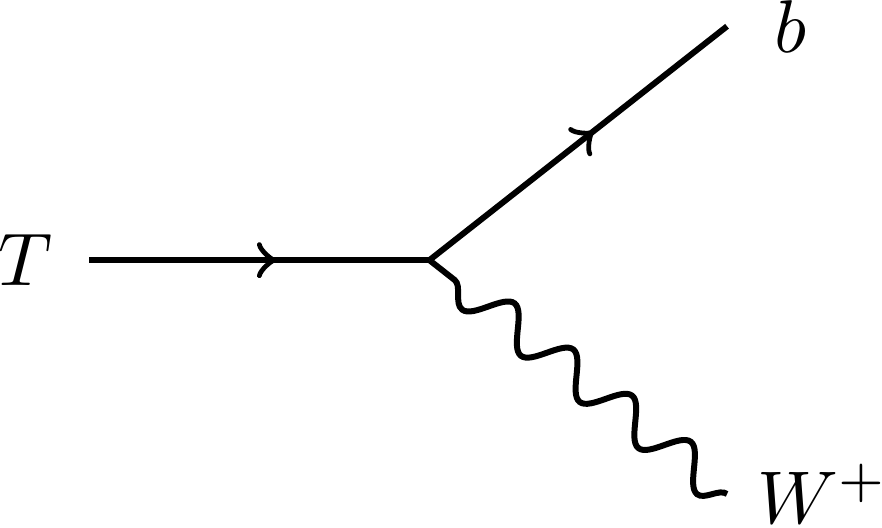}\label{fig:TbW}}
\subfigure[]{\includegraphics[width=0.28\textwidth,clip]{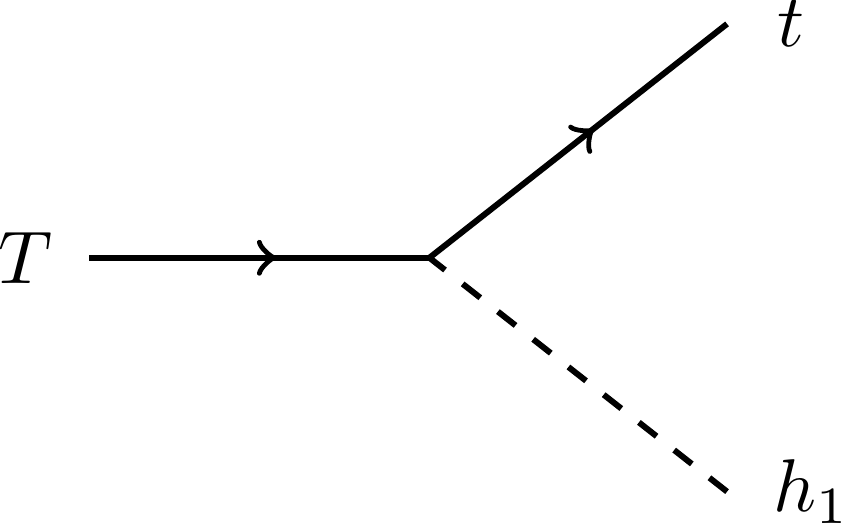}\label{fig:Tth1}}
\subfigure[]{\includegraphics[width=0.28\textwidth,clip]{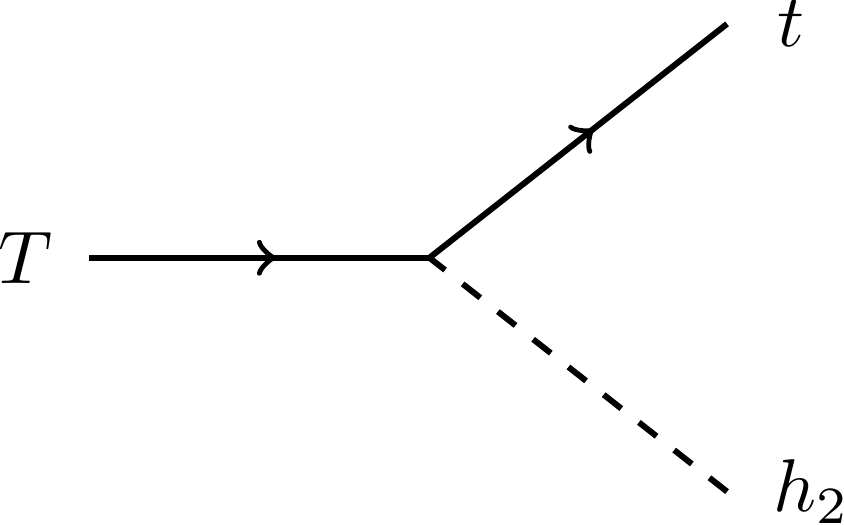}\label{fig:Tth2}}
\subfigure[]{\includegraphics[width=0.28\textwidth,clip]{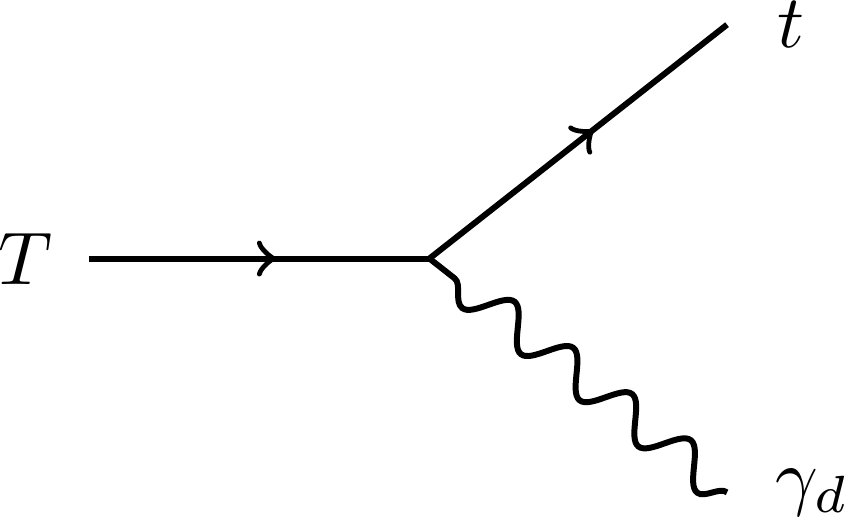}\label{fig:TtDark}}
\end{center}
\caption{\label{fig:TdecayAG} Representative Feynman diagrams for VLQ decays into (a-c) $Zt$, $Wb$, $h_1 t$. Since $T$ is charged under both the SM and $U(1)_d$, the $T$ is allowed to decay into (d,e) $h_2$ and $\gamma_d$.} 
\end{figure}

Traditionally, searches for the VLQ rely on the $T \rightarrow t Z$, $T \rightarrow b W$, and $T \rightarrow t h_1$ decays, as shown in Fig~\ref{fig:TdecayAG}. However, in the scenario where $T$ is charged under both the SM and $U(1)_d$, new decay modes into the $T \rightarrow t h_2$ and $T \rightarrow t \gamma_d$ appear, which alters $T$ phenomenology significantly. Partial widths into $Z/W/h_1$ in the limit $|\varepsilon|,|\theta_L^t|, |\theta_S|   \ll1$ and $v_{EW},\,v_d\ll M_T$ are\footnote{To produce numerical results and plots, however, we will use exact width expressions.}

\begin{figure}[!htb!]
\begin{center}
\subfigure[]{\includegraphics[width=0.49\textwidth,clip]{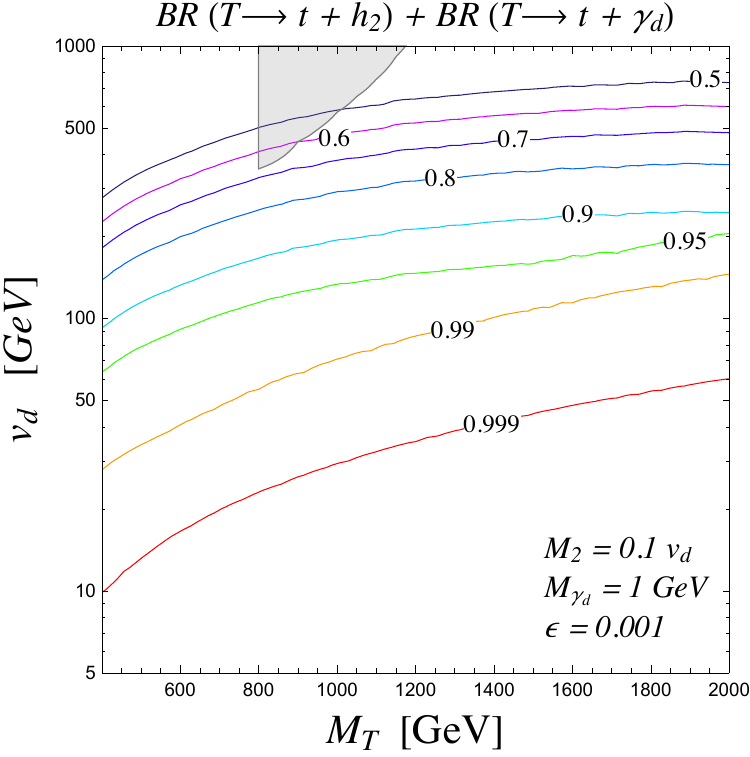}\label{fig:BrSorAD1}}
\subfigure[]{\includegraphics[width=0.49\textwidth,clip]{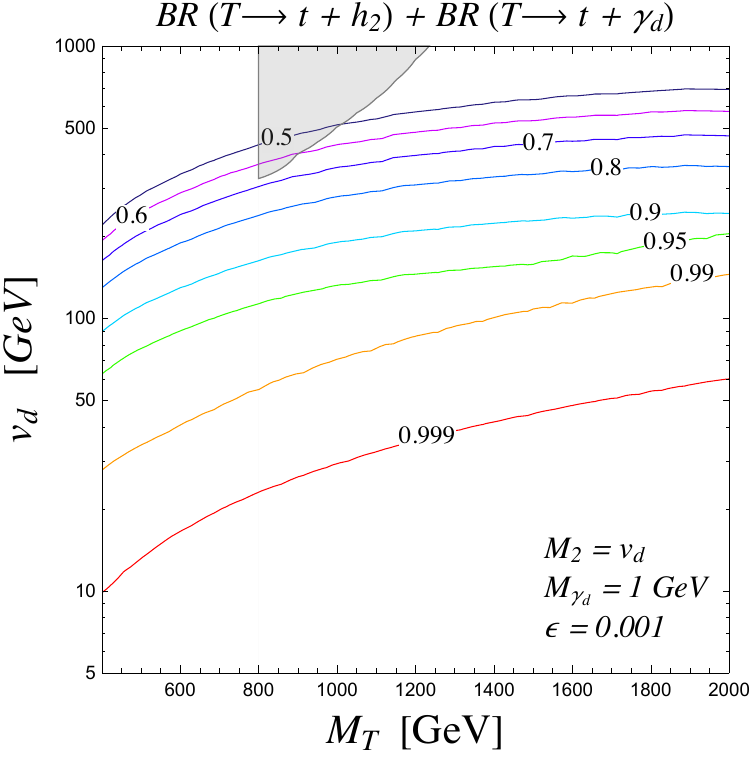}\label{fig:BrSorAD2}}
\subfigure[]{\includegraphics[width=0.49\textwidth,clip]{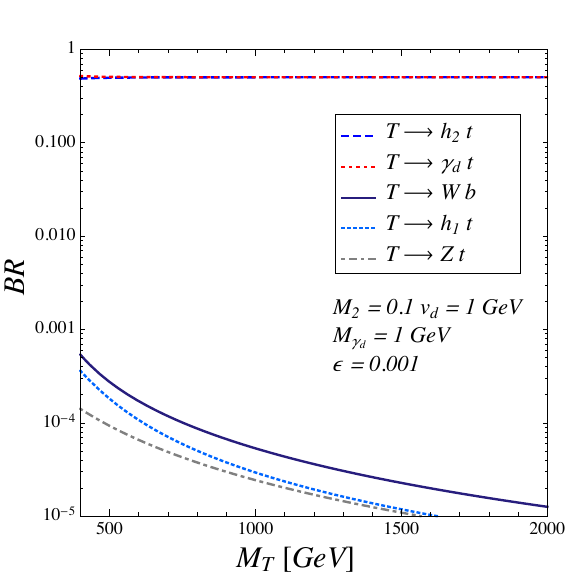}\label{fig:BrT1}}
\subfigure[]{\includegraphics[width=0.49\textwidth,clip]{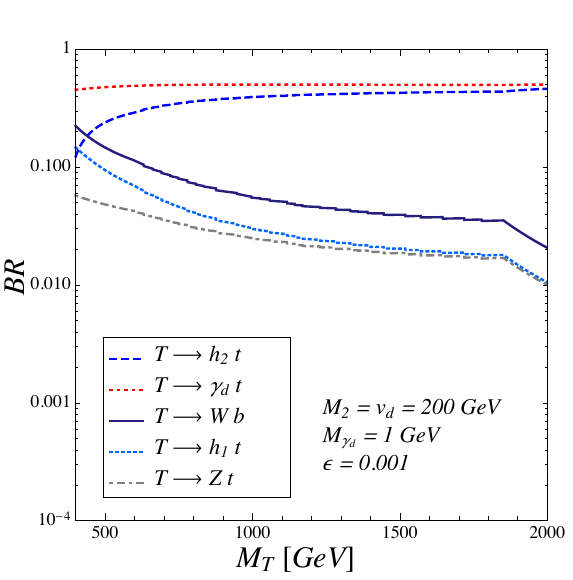}\label{fig:BrT2}}
\end{center}
\vspace{-5 mm}\caption{\label{fig:TBRSorAD}(a,b)The combined branching ratio $\text{BR}(T\rightarrow t+h_2) + \text{BR}(T\rightarrow t+\gamma_d)$ in $M_T - v_d$ plane for maximally allowed $\sin \theta^t_L$ and $\sin \theta_S$ in Fig.~\ref{thetaL-MT-plane} and Eq.~(\ref{eq:sthlim}), respectively.  In (a)   $M_2 = 0.1 v_d$ and (b) $M_2 = v_d$. The full branching ratios of the $T$ as a function of $M_T$ for (c) $M_{2} = 0.1 v_d = 1$ GeV and (d) $M_{2} =  v_d = 200$ GeV. For all subfigures we assume $\varepsilon= 0.001$. 
} 
\end{figure}
\begin{figure}[!htb!]
\begin{center}
\subfigure[]{\includegraphics[width=0.49\textwidth,clip]{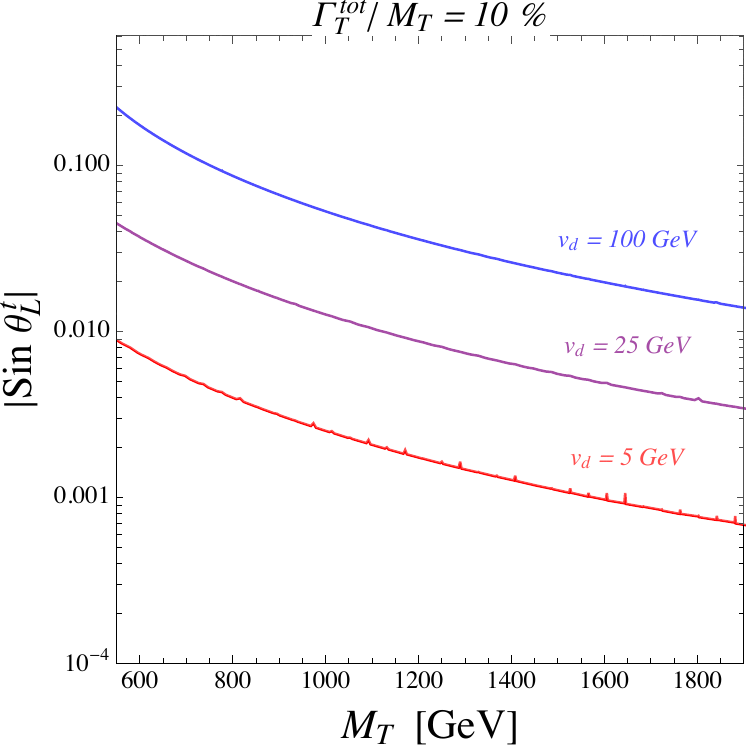}\label{fig:WidthT1}}
\subfigure[]{\includegraphics[width=0.49\textwidth,clip]{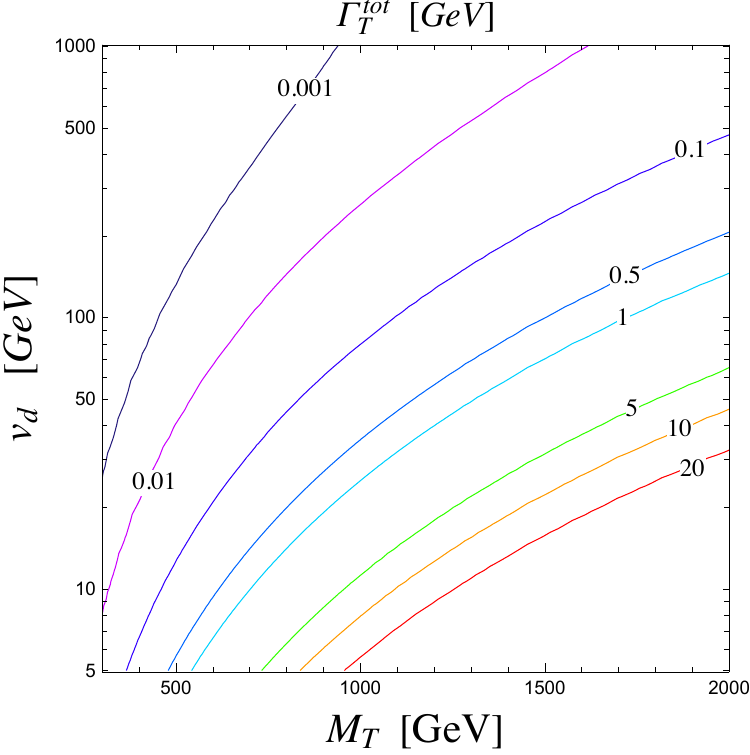}\label{fig:WidthT2}}
\end{center}
\vspace{-5 mm}\caption{\label{fig:Twidth} (a) Contours of $\Gamma_T^{tot}/M_T=0.1$ in the $M_T-\sin\theta_L^t$ plane for various $v_d$.  (b) The total width ($\Gamma^{\text{tot}}_T$) of $T$ in $M_T - v_d$ plane for $\sin \theta^t_L =  0.001$. For both sub-figures, we set $\varepsilon = 0.001$, $M_{\gamma_d} = M_2 = 2$ GeV, and $\sin\theta_S$ to its maximal value in Eq.~(\ref{eq:sthlim}).
} 
\end{figure}
\begin{eqnarray}
\Gamma(T\rightarrow t\,Z)&\approx&\Gamma(T\rightarrow t\,h_1)\approx\frac{1}{2}\Gamma(T\rightarrow b\,W)\approx\frac{1}{32\pi}\frac{M_T^3}{v^2_{EW}}\sin^2\theta_L^t~.\label{eq:TtoSM}
\end{eqnarray}
For large $M_T$, the partial widths of $T$ into fully SM final states are proportional to $\sim \sin^2\theta^t_L\,M_T^3/v^2_{EW}$ due to the Goldstone equivalence theorem. The partial widths into $h_2$ and $\gamma_d$ in the limit $|\varepsilon|,|\theta_L^t|, |\theta_S|   \ll1$ and $v_d,\,v_{EW}\ll M_T$ are

\begin{eqnarray}
\Gamma(T\rightarrow t\,\gamma_d)&\approx&\Gamma(T\rightarrow t\,h_2)\approx \frac{1}{32\pi}\frac{M_T^5}{M_t^2v_d^2}\frac{\sin^2\theta_L^t}{\left(1+(M_T/M_t)^2\sin^2\theta_L^t\right)^2}~.\label{eq:TtoDark}
\end{eqnarray}
Hence, the ratios of the rates of VLQ decays into the dark Higgs/photon and into fully SM final states are
\begin{eqnarray}
\frac{\Gamma(T\rightarrow t+h_2/\gamma_d)}{\Gamma(T\rightarrow t/b+W/Z/h_1)}\sim \left(\frac{M_T}{M_t}\right)^2\left(\frac{v_{EW}}{v_d}\right)^2\frac{1}{\left({1+(M_T/M_t)^2\sin^2\theta_L^t}\right)^2}~.\label{eq:enhancement}
\end{eqnarray}
There are two enhancements: (1) the $(v_{EW}/v_d)^2$ enhancement since decays into longitudinal dark photons are enhanced by $v_d^{-2}$ compared to decays into longitudinal SM bosons which are proportional to $v^{-2}_{EW}$. (2) If $|\sin\theta_L^t|\lesssim M_t/M_T$ there is a $(M_T/M_t)^2$ enhancement since the right-handed top-VLQ mixing angle is larger than left-handed mixing due to a fermion mass hierarchy as seen in Eqs.~(\ref{eq:thetaRlim}-\ref{eq:Gauge_Top}).  However, note that for fixed $|\sin\theta_L^t|\gg M_t/M_T$, the fermion mass hierarchy enhancement cancels and only the $(v_{EW}/v_d)^2M_t^2/M_T^2\sin^4\theta_L^t$ survives.  Although, note there is now a suppression from the fermion mass hierarchy and $|\sin\theta_L^t|\gg M_t/M_T$ is disfavored by perturbative unitarity [Eq.~(\ref{eq:sthLtBND})].  This is because in this limit $|\sin\theta_R^t|\rightarrow 1$ and does not grow with $M_T$.

Equation~(\ref{eq:enhancement}) shows that even in the absence of a fermionic mass hierarchy ($M_T\sim M_t$), $T$ decays into light dark sector bosons are still strongly enhanced. This can be clearly seen in Figure \ref{fig:TBRSorAD}(a,b) where we show contours of the total VLQ branching ratio in $h_2$ and $\gamma_d$.  Note that $\text{BR}(T\rightarrow t+h_2) + \text{BR}(T\rightarrow t+\gamma_d) \sim 0.99$ for $v_d\lesssim 30$~GeV in the entire $M_T$ range. As $M_T$ increases, branching ratios into the dark photon/Higgs increase due to the fermionic mass hierarchy, as discussed above.  Fig.~\ref{fig:BrSorAD2} is the same as Fig.~\ref{fig:BrSorAD1} with a different choice of $M_2$.  The results in both Fig.~\ref{fig:TBRSorAD}(a,b) are very similar, showing the conclusions about the branching ratio dependence on boson and fermion mass hierarchies are robust against model parameters.   The reach of current searches into $th,\,tZ,\,$ and $bW$~\cite{Aaboud:2018pii} are shown in the gray shaded regions.  We have rescaled the results of Ref.~\cite{Aaboud:2018pii} according to the branching ratios in our model.  There were no limits below $M_T=800$~GeV, hence the exlusion region is truncated.  As can be seen, the traditional searches are largely insensitive to our model and our approach provides a new avenue to search for $T$.  New search strategies are necessary depending on the decays of $\gamma_d,\,h_2$ as we will discuss in section \ref{sec:collider}.

In Fig.~\ref{fig:TBRSorAD}(c,d) we show the branching ratios of $T$ into all final states, including $W,\,Z,\,h_1$.  The $T$ branching ratios into the fully SM particles are less than $\sim 1\%$ for smaller $M_2=0.1\,v_d=1$~GeV as shown in Fig.~\ref{fig:BrT1}.  For enhanced dark sector mass scales $M_2=v_d=200$~GeV the rates to the SM final states can reach at most $\sim 45\%$ for $M_T\sim300$~GeV shown in Fig.~\ref{fig:BrT2}, but then fall to the percent level for higher VLQ masses.

There is a kink in Fig.~\ref{fig:BrT2} around $M_T\sim 1.9$~TeV.  For $M_T\lesssim 1.9$~TeV EW precision constraints on $\sin\theta_L^t$ are the most stringent and for $M_T\gtrsim~1.9$~TeV the perturbativity bounds on $\lambda_t$ are most constraining [see Fig.~\ref{thetaL-MT-plane}].  The EW precision and perturbativity bounds on $\sin\theta_L^t$ have different dependendencies on $M_T$, hence the kink.  The branching ratios into $W/Z/$Higgs become flat for $M_T$ approaching 1.9~TeV. Without perturbative unitarity constraints, these branching ratios would eventually increase due to the suppression of $T\rightarrow t\,h_d/\gamma_d$ by the fermionic mass hierarchy for the limit of $M_T\gg M_t$ and fixed $\sin\theta_L^t$, as discussed around Eq.~(\ref{eq:enhancement}). Perturbative unitarity instead prevents this increase from occurring. Once perturbativity constraints are dominant $\sin\theta_L^t\sim M_T/M_t$, the fermion mass hierarchy enhancement reasserts itself, and branching ratios into fully SM final states decrease precipitously.

Finally, in the limit $M_t\ll M_T$ and $v_d \ll v_{EW}$, the total width of the VLQ normalized to $M_T$ is
\begin{eqnarray}
\frac{\Gamma_T^{tot}}{M_T}=\frac{\Gamma(T\rightarrow t\,\gamma_d)+\Gamma(T\rightarrow t\,h_2)}{M_T}\approx  \frac{1}{16\pi}\frac{M_T^4}{M_t^2\,v_d^2}\frac{\sin^2\theta_L^t}{\left(1+(M_T/M_t)^2\sin^2\theta_L^t\right)^2}~.\label{eq:GamTtot}
\end{eqnarray}
Due to the very large enhancement of $M_T^4/M_t^2/v_d^2$, the mixing angle $\sin\theta_L^t$ must be quite small for $T$ to be narrow.  In Fig.~\ref{fig:WidthT1} we show contours of fixed $\Gamma_T^{tot}/M_T=10\%$ in the $\sin\theta_L^t-M_T$ plane for various dark Higgs vevs $v_d$.  When compared to the constraints in Fig.~\ref{thetaL-MT-plane}, it is clear that the constraint $T$ be narrow with $\Gamma_T^{tot}\lesssim 10\%\,M_T$ is by far the strongest constraint on $\sin\theta_L^t$.  In Fig.~\ref{fig:WidthT2} we show the total width $\Gamma_T^{tot}$ in the $v_d-M_T$ plane.  As is clear, VLQ total width grows for small $v_d$ and larger $M_T$.

%%%%%%%%%%%%%%%%%%%%%%%%%%%%%%%%%%%%%%%%%%%%%%%%%%%%%%
\section{Decay of the dark photon} 
\label{sec:Adecay}
%%%%%%%%%%%%%%%%%%%%%%%%%%%%%%%%%%%%%%%%%%%%%%%%%%%%%%
%

We now discuss the dark photon $\gamma_d$ decays, since this specifies experimental signatures in the VLQ decay $T \rightarrow t \gamma_d$. The lowest order (LO) $\gamma_d$ partial decay widths can be computed using the couplings to the light fermions from the covariant derivative in Eq.~(\ref{eq:NewCovar}).  However, this does not take into account the higher-order QCD corrections and hadronic resonances. To reflect these combined effects, we follow Ref.~\cite{Curtin:2014cca} and utilize the experimental data on electron positron collisions \cite{Tanabashi:2018oca}
\bea
\label{eq:Rs} 
R(M_{\gamma_d}) \equiv \frac{\sigma(e^+ e^- \rightarrow hadrons)}{\sigma(e^+ e^- \rightarrow \mu^+ \mu^-)}~.
\eea
Since $\gamma_d$ couplings are approximately electromagnetic, hadronic decays of $\gamma_d$ can be incorporated into the total width of $\gamma_d$ via 
\bea
\label{eq:TotWidthAD} 
\Gamma^{tot}_{\gamma_d} &=& R(M_{\gamma_d}) \Gamma (\gamma_d \rightarrow \mu^+ \mu^-) + \sum_{f = e, \mu, \tau, \nu_e, \nu_{\mu}, \nu_{\tau} } \Gamma (\gamma_d \rightarrow f \overline{f})\nonumber\\
&\approx&\frac{\varepsilon^2\,e^2}{12\,\pi}M_{\gamma_d}\left[R(M_{\gamma_d})+\sum_{\ell=e,\mu\tau}\theta(M_{\gamma_d}-2\,M_\ell)\right]~.
\eea
We have used the approximation $\varepsilon\ll1$ and $M_{\gamma_d}\ll M_Z$ as in Eq.~(\ref{eq:NewCovar}).  We have also assumed there are no DM candidates with mass $2\,M_{DM}<M_{\gamma_d}$ and that $2\,M_2>M_{\gamma_d}$ so that $\gamma_d\rightarrow$DM and $\gamma_d\rightarrow 2\,h_2$ decays are forbidden.
\begin{figure}[t]
\begin{center}
\subfigure[]{\includegraphics[width=0.5\textwidth,clip]{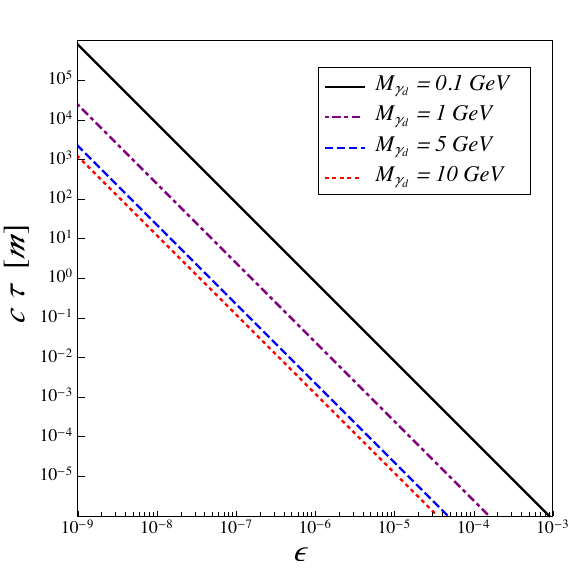}\label{fig:Width_AD}}
\subfigure[]{\includegraphics[width=0.475\textwidth,clip]{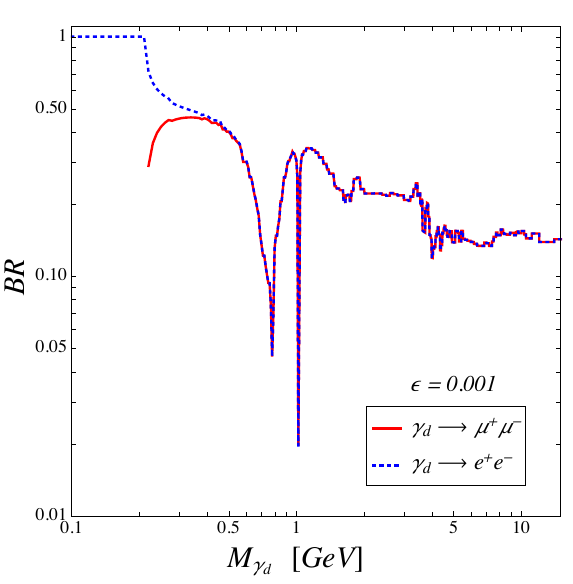}\label{fig:BRAD}}
\end{center}
\vspace{-5 mm}\caption{ (a) Decay length of $\gamma_d$ as a function of the kinetic mixing parameter $\varepsilon$ for various $M_{\gamma_d}$. (b) Branching ratios of $\gamma_d$ into (solid red) $\mu^+\mu^-$ and (dotted blue) $e^+e^-$ as a function of $M_{\gamma_d}$ for $\varepsilon=0.001$.
} 
\end{figure}

The lifetime of the dark photon can be calculated by
\begin{eqnarray}
\tau = \frac{1}{\Gamma^{tot}_{\gamma_d}}~.
\end{eqnarray}
Hence, the $\gamma_d$ lifetime is inversely proportional to $\varepsilon^2$.  For small kinetic mixing parameter the dark photon can be quite long lived and have a large decay length.  In Fig.~\ref{fig:Width_AD} we show the decay length $c\tau$ of the dark photon as a function of the kinetic mixing parameter $\varepsilon$ for various dark photon masses.  For $\varepsilon$ in the range of $1-5\times 10^{-6}$ the decay length can be $c\tau\sim 1$~mm.  As discussed in the next section, this can lead to a spectacular collider signature of displaced vertices.

In Fig.~\ref{fig:BRAD} we show the branching ratios of the dark photon into electrons and muons.  This reproduces the results from Ref.~\cite{Curtin:2014cca}, which we have recalculated and included for completeness.  The branching ratios of the dark photon into electrons and muons are almost identical when $M_{\gamma_d} > 2 M_{\mu}$. For much lower masses below $\sim 200$ MeV, the $\gamma_d$ decay to muons is kinematically closed, and hence $\gamma_d \rightarrow e^+ e^-$ decays dominate.  The multiple dips in the branching ratios starting around $M_{\gamma_d}\sim 770$~MeV are attributed to hadronic resonances $\rho$, $\omega$, $\phi$, $\rho'$, $J/\psi$, $\psi (2S)$, and $\Upsilon(nS)$ for $n=1,2,3,4$~\cite{Tanabashi:2018oca}.

%%%%%%%%%%%%%%%%%%%%%%%%%%%%%%%%%%%%%%%%%%%%%%%%%%%%%%
\section{Searching for the dark photon with $T \rightarrow t \gamma_d$ decays}
\label{sec:collider}
%%%%%%%%%%%%%%%%%%%%%%%%%%%%%%%%%%%%%%%%%%%%%%%%%%%%%%
We now discuss the collider signatures of this model.  As discussed previously, the pair production of $T$ only depends on the spin and mass of $T$ and ${\rm BR}(T\rightarrow t\,\gamma_d)\approx 50\%$ in a very large range of parameter space.  Hence, the production rate of the dark photon is at QCD rates and largely independent of the model parameters.  The major model dependence comes from the lifetime of $\gamma_d$.  If $\varepsilon$ is sufficiently small, the dark photon becomes long-lived. 

The decay length of the dark photon from $T$ decays is
\begin{eqnarray}
d = \bar{b} c \tau~,
\end{eqnarray}
where $c \tau$ is a proper lifetime as shown in Fig.~\ref{fig:Width_AD} and $\bar{b}$ is the average boost of the dark photon.  Assuming the VLQs are produced mostly at rest, the boost is 
\begin{eqnarray}
\bar{b}  &=& \frac{|\overrightarrow{p}_{\gamma_d}|}{M_{\gamma_d}}=\frac{1}{2 M_{\gamma_d} M_T} \sqrt{ (M^2_T - M^2_{\gamma_d} - M^2_t )^2 - 4 M^2_{\gamma_d} M^2_t} \\ \nonumber
&\xrightarrow[M_T\gg M_{\gamma_d},M_t]{}&\frac{M_T}{2 M_{\gamma_d}}~,
\end{eqnarray}
where $|\overrightarrow{p}_{\gamma_d}|$ is the dark photon 3-momentum. Using the total $\gamma_d$ width in Eq.~(\ref{eq:TotWidthAD}), we can then solve for the decay length:
\begin{eqnarray}
d=580~\mu{\rm m}\times \frac{7}{R(M_{\gamma_d})+\sum_{\ell=e,\mu\tau}\theta(M_{\gamma_d}-2\,M_\ell)}\left(\frac{M_T}{1~{\rm TeV}}\right)\left(\frac{1~{\rm GeV}}{M_{\gamma_d}}\right)^2\left(\frac{10^{-4}}{\varepsilon}\right)^2~.
\end{eqnarray}
Hence, for reasonable parameter choices, the decay length of the dark photon can be several hundreds of microns.  The precise direction of the dark photon in the detector will determine if it appears as a displaced vertex or where it will decay in the detector. Nevertheless, for $d\lesssim 500~\mu$m the dark photon decay can be considered prompt, for $d=1~{\rm mm}-1~{\rm m}$ it will be a displaced vertex, for $d\sim1~{\rm m}-10~{\rm m}$ the dark photon will decay in the detector, and $d\gtrsim 10~{\rm m}$ the dark photon will decay outside the detector~\cite{Kim:2018mks}.  Hence, we can solve for the values of $\varepsilon$ for these various scenarios:
\begin{eqnarray}
\varepsilon&=&\left(\frac{7}{R(M_{\gamma_d})+\sum_{\ell=e,\mu\tau}\theta(M_{\gamma_d}-2\,M_\ell)}\right)^{1/2}\left(\frac{M_T}{1~{\rm TeV}}\right)^{1/2}\left(\frac{1~{\rm GeV}}{M_{\gamma_d}}\right)\\
&&\quad\quad\times \begin{cases}\gtrsim 1\times10^{-3}&\text{for prompt decays}\\
2.4\times10^{-6}-7.6\times10^{-5}&\text{for displaced vertices}\\
7.6\times10^{-7}-2.4\times10^{-6}&\text{for decays in detector}\\
\lesssim 7.6\times10^{-7}&\text{for decays outside the detector~.}\end{cases}\nonumber
\end{eqnarray}

If the dark photon decays outside the detector it is unobserved, giving rise to the final state characterized by $t \bar t + \slashed{E}_T$.  This is the same signature as pair produced scalar tops, $\widetilde{t}$, in R-Parity conserving SUSY models with the decays $\widetilde{t}\rightarrow t\,\widetilde{\chi}^0_1$, where $\widetilde{\chi}^0_1$ is the lightest superpartner and stable.  Hence,  the currently available CMS \cite{Sirunyan:2017xse,Sirunyan:2017leh,Sirunyan:2017cwe,Sirunyan:2017kqq,Sirunyan:2017pjw,CMS-PAS-SUS-19-005} and ATLAS \cite{Aaboud:2017ayj,Aaboud:2017nfd} searches for stop pair production can be used to obtain constraints on the model presented here. In the limit of large gluino/squark masses, the most stringent bound is at 13 TeV excludes stop masses up to 1225 GeV for a massless $\tilde{\chi}^0_1$~\cite{CMS-PAS-SUS-19-005}.  Since Ref.~\cite{CMS-PAS-SUS-19-005} assumes ${\rm BR}(\widetilde{t}\rightarrow t\,\widetilde{\chi}^0_1)=1$, the corresponding $95 \%$ CL upper limit on the NLL-NLO stop pair production cross section is given by $\sim 1.3$~fb~\cite{Borschensky:2014cia}. 

\begin{figure}[t]
\begin{center}
\subfigure[]{\includegraphics[width=0.49\textwidth,clip]{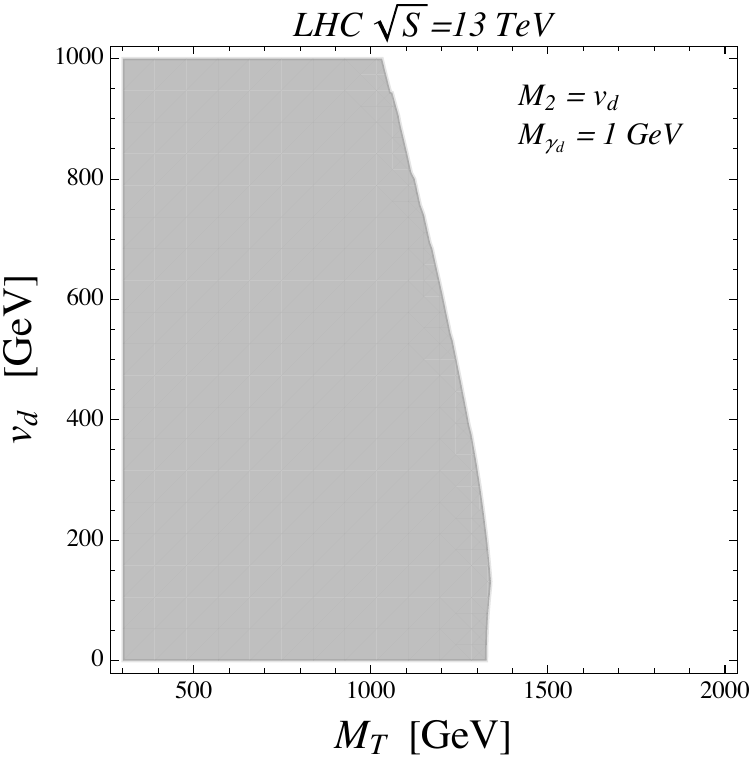}\label{fig:Bounds_Stop}}
\subfigure[]{\includegraphics[width=0.49\textwidth,clip]{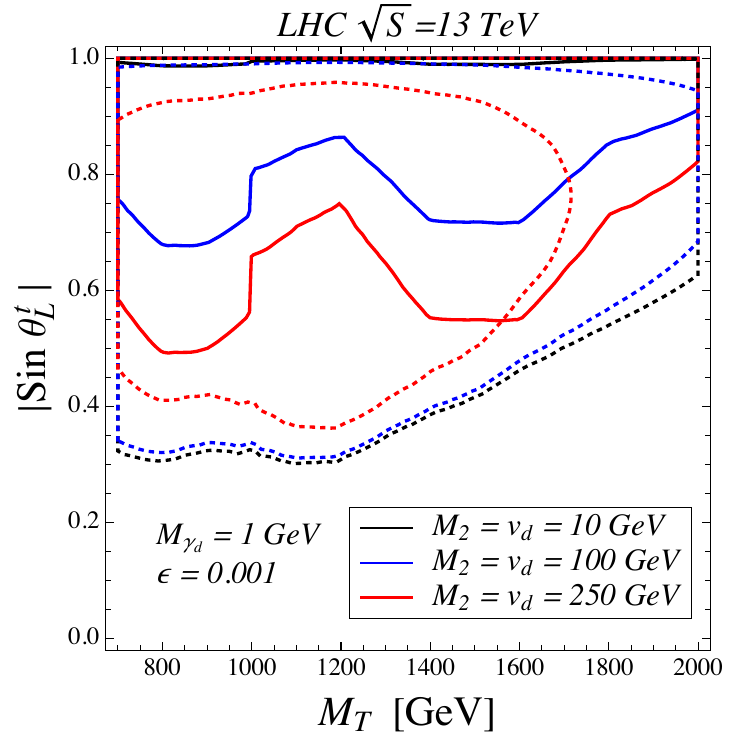}\label{fig:Bounds_Ttz3}}
\end{center}
\vspace{-5 mm}\caption{\label{fig:Bounds} (a) $95 \%$ CL exclusion regions in $M_T-v_d$ plane from the VLQ pair production with the $T \rightarrow t+  \gamma_d/\nu\bar{\nu}$ decays, assuming $\gamma_d$ is long lived. The constraint is obtained from re-interpreting the bounds from the CMS \cite{CMS-PAS-SUS-19-005} stop searches at 13 TeV. (b) $95 \%$ CL exclusion regions in $M_T-\sin\theta_L^t$ plane based on the single production of the VLQ. The dotted constraint is obtained from a simple recast of the ATLAS \cite{Aaboud:2018zpr} results on the single production of $T$ with the decay $T \rightarrow t Z (\rightarrow \nu \bar{\nu})$. The solid lines are taken from a recasting of a CMS \cite{CMScollabSearch:2019} search for single production of $T$ with fully hadronic decays into Higgs or $Z$. 
} 
\end{figure}

Since both stop and $T$ pair production yield similar kinematic distributions in the final states, the efficiencies of two searches are quite similar \cite{Kraml:2016eti}. The upper bound on the stop pair production cross section can then be reinterpreted as a bound on the VLQ pair production cross section:
\begin{eqnarray}
\sigma ( p p \rightarrow T \overline{T} ) \times \left(\text{BR}(T \rightarrow t \gamma_d)+\text{BR}(T\rightarrow t Z\rightarrow t \nu\bar{\nu})\right)^2\leq 1.3~{\rm fb}~.
\end{eqnarray}
In Fig.~\ref{fig:Bounds_Stop} we show this limit in the $v_d-M_T$ plane for a dark photon mass of $M_{\gamma_d}=1$~GeV (gray region is ruled out). We used the $T$ branching ratios in Fig.~\ref{fig:TBRSorAD} and the NNLO $T\overline{T}$ cross section in Fig.~\ref{fig:T_prod_13}.  As shown in Sec.~\ref{sec:proddec}, the production and decay rates of the VLQ are relatively independent of model parameters and this result is robust. We find that VLQ masses
\begin{eqnarray}
M_T \lesssim 1.2 \; \text{TeV}~,
\end{eqnarray}
are excluded for $M_{\gamma_d} = 1$ GeV and $v_d \lesssim 500$ GeV when the dark photon is stable on collider time scales. The bound can be weakened for higher values of $v_d$ since the branching ratio of $T$ into SM bosons with visible decays increases, suppressing $\text{BR}( T \rightarrow t \gamma_d)$ as displayed in Figure \ref{fig:TBRSorAD}. 

\begin{figure}[!htb!] 
\begin{center}
\subfigure[]{\includegraphics[width=0.49\textwidth,clip]{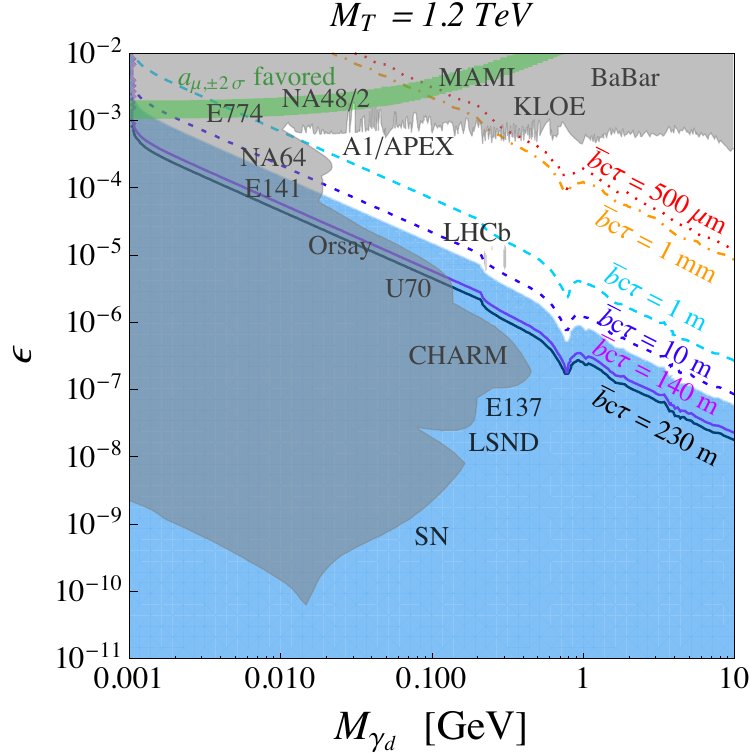}\label{fig:Final_1d2TeV}}
\subfigure[]{\includegraphics[width=0.49\textwidth,clip]{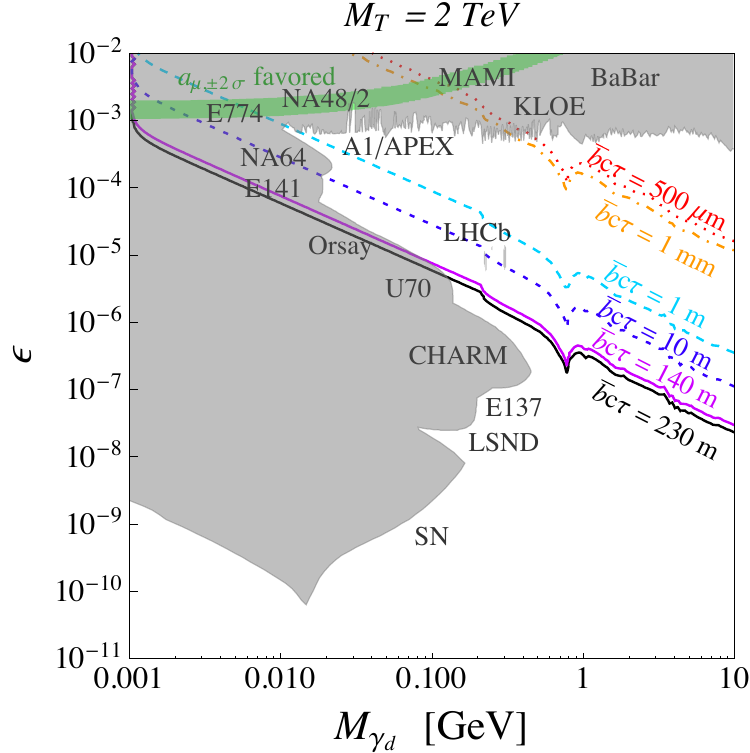}\label{fig:Final_2TeV}}
\end{center}
\vspace{-5 mm}\caption{\label{fig:Final} Various decay lengths of the dark photon originating from VLQs with masses (a) 1.2 TeV and (b) 2 TeV in $\varepsilon-M_{\gamma_d}$ plane.  The blue shaded regions are excluded by searches for stop pair production with decay $\widetilde{t}\rightarrow t\,\widetilde{\chi}^0_1$.  Gray shaded regions correspond to existing $90\%$ confidence level limits from the SLAC and Fermilab beam dump experiments E137, E141, and E774 \cite{Bjorken:2009mm,Bjorken:1988as,Riordan:1987aw,Bross:1989mp}, the U70 accelerator \cite{Blumlein:2013cua, Blumlein:2011mv}, LHCb \cite{Aaij:2017rft, Ilten:2018crw}, NA64 \cite{Banerjee:2018vgk}, the electron and muon anomalous magnetic moment $a_{\mu}$ \cite{Pospelov:2008zw,Davoudiasl:2012ig,Endo:2012hp}, KLOE \cite{Babusci:2012cr, Archilli:2011zc}, WASA-at-COSY \cite{Adlarson:2013eza}, the test run results reported by APEX \cite{Abrahamyan:2011gv} and MAMI \cite{Merkel:2011ze}, an estimate using a BaBar result \cite{Bjorken:2009mm,Reece:2009un,Aubert:2009cp}, and a constraint from supernova cooling \cite{Chang:2016ntp,Chang:2018rso}. The shaded green regions are favored to explain the muon anomalous magnetic moment \cite{Pospelov:2008zw} at $95\%$ confidence level.
} 
\end{figure}
Searches for single $T$ production can be important if $t-T$ mixing is not too small.  It is clear from Figure \ref{fig:T_prod_13} that for $\sin \theta^t_L \sim 0.1$ the single production dominates over the pair production at high VLQ masses. Refs. \cite{Backovic:2015lfa,Backovic:2015bca} showed that the $T \rightarrow t Z (\rightarrow \nu \bar{\nu})$ channel displays a superior performance in prospects for discovering the $T$.  The signature is then $T\rightarrow t+\slashed{E}_T$, which is the same as for $T\rightarrow t\gamma_d$ when $\gamma_d$ is long lived.  The ATLAS collaboration~\cite{Aaboud:2018zpr} presented results on the single production of $T$ with the decay $T \rightarrow t Z (\rightarrow \nu \bar{\nu})$. Assuming that efficiencies of $T\rightarrow tZ\rightarrow t\nu\bar{\nu}$ and $T\rightarrow t\gamma_d$ searches are the same, we re-interpret the $95\%$ CL upper limit on the cross section in Ref. \cite{Aaboud:2018zpr} to derive constraints on $M_T - \sin \theta^t_L$ plane, as shown as dotted lines in Figure \ref{fig:Bounds_Ttz3}.   The regions within the curves are ruled out.  As in Fig.~\ref{fig:Bounds_Stop}, we consider both $T\rightarrow t\gamma_d$ where the dark photon is assumed to escape the detector, and $T\rightarrow t\nu\bar{\nu}$. For VLQ masses around $M_T = 1 \sim 2$ TeV, the limits on $\sin \theta^t_L$ are
\begin{eqnarray}
|\sin \theta^t_L| \lesssim 0.3 \sim 0.6 ~,
\end{eqnarray}
where the stronger bounds are expected for smaller values of $v_d$ due to the enhancement of the branching ratio $\text{BR}(T \rightarrow t \gamma_d)$.  For smaller $|\sin\theta_L^t|$ the single VLQ production rate is too small to be detectable yet.  For larger $\sin\theta_L^t$ and larger $v_d$, the VLQ essentially decays like a top quark with a near 100\% branching ratio into $Wb$.  Hence, the branching ratio to $t\gamma_d$ is suppressed and a gap appears for $|\sin\theta_L^t|\gtrsim 0.9$ and $v_d=250$~GeV.  These bounds are, however, weaker as compared to the EW precision test [see Figure \ref{thetaL-MT-plane}].

CMS has also performed a recent search for electroweak production of T decaying through Z and Higgs channels with fully hadronic decays \cite{CMScollabSearch:2019}. These searches can be re-interpretted into constraints in the $M_T - \sin \theta^t_L$ plane. The $95$\% CL exclusions are shown as solid lines in Figure \ref{fig:Bounds_Ttz3} with the regions inside the curves ruled out. In the small $v_d$ limit ($v_d=10$~GeV) the branching ratios into SM final states are neglible so there are not strong constraints. As $v_d$ increases the branching ratios become viable and some constraints emerge. In the high $v_d$ the generic search constraints start to become more stringent than $T \rightarrow t Z (\rightarrow \nu \bar{\nu})$ constraints.  Hence, there is a complementarity between the fully hadronic and missing energy searches.

Figure \ref{fig:Final} shows the decay lengths of dark photons originating from the VLQ  with masses (a) $M_T=1.2$ TeV and (b) $M_T=2$ TeV in $M_{\gamma_d}-\varepsilon$ plane. We show several lines of the dark photon decay length $d=\overline{b}c\tau$ that are indicative of prompt decays ($d$=500 $\mu$m), displaced vertices ($d$=1~mm), decays in the detector ($d$=1 m), and decays outside the detector ($d$=10 m).  Additionally, there is a proposed MATHUSLA detector~\cite{Curtin:2017izq} to search for long lived particles.  MATHUSLA will be on the surface $140-230$ m away from the interaction point.  Hence, we also show lines for dark photons that could decay inside the MATHUSLA detector.  The blue shaded regions are excluded by searches for stop pair production with decay $\widetilde{t}\rightarrow t\,\widetilde{\chi}^0_1$, as discussed above.  This blue exclusion region exists for $M_T\lesssim 1.3$~TeV. Hence, it appears in Fig.~\ref{fig:Final_1d2TeV} but not Fig.~\ref{fig:Final_2TeV}. The grey shaded regions are excluded by various low energy experiments~\cite{Battaglieri:2017aum} and supernova measurements~\cite{Chang:2016ntp,Chang:2018rso}.   As can be clearly seen, searches for $T\rightarrow t\,\gamma_d$ with a wide range of possible signals can cover a substantial portion of the parameter space.  This is because in the model presented here the production of $\gamma_d$ from VLQ production is largely independent of the small kinetic mixing parameter.  Hence, the production rate of $\gamma_d$ is unsuppressed at low $\varepsilon$ and the LHC can be quite sensitive to this region.

The dark photon branching ratios into $e^-e^+$ and $\mu^-\mu^+$ is non-negligible as shown in Fig.~\ref{fig:BRAD}.  Hence, the most promising signature of the $T\rightarrow t\,\gamma_d$ would be the leptonic decays of the dark photon, which would help avoid large QCD backgrounds.  Since the dark photon is highly boosted, its decay products are highly collimated.  The angular distance between the leptons from $\gamma_d$ decays can be estimated as
\begin{eqnarray}
\Delta R_{\ell\ell}\sim \frac{2\,M_{\gamma_d}}{E_{\gamma_d}}=\frac{4\,M_{\gamma_d}}{M_T}=4\times 10^{-3}\left(\frac{M_{\gamma_d}}{1~{\rm GeV}}\right)\left(\frac{1~{\rm TeV}}{M_T}\right)~,
\end{eqnarray}
where $\Delta R_{\ell\ell}=\sqrt{(\phi_{\ell^-}-\phi_{\ell^+})^2+(\eta_{\ell^-}-\eta_{\ell^+})^2}$, $\phi$ are the azimuthal angles of the leptons, and $\eta$ are their rapidities.  At such small angular separation, the leptons are very difficult to isolate and the dark photon can give rise to so-called ``lepton jets''~\cite{ArkaniHamed:2008qp,Rizzo:2018vlb} which are highly collimated clusters of electrons and muons.  In fact, for not too small kinetic mixing $\varepsilon$, there could be displaced lepton jets or even lepton jets originating in the detector.

%%%%%%%%%%%%%%%%%%%%%%%%%%%%%%%%%%%%%%%%%%%%%%%%%%%%%%
\section{Conclusion}
\label{sec:conc}
%%%%%%%%%%%%%%%%%%%%%%%%%%%%%%%%%%%%%%%%%%%%%%%%%%%%%%
%
In this paper we have studied a model with an up-type VLQ charged under a new $U(1)_d$, where the $U(1)_d$ gauge boson kinetically mixes with the SM hypercharge.   One of the most significant aspects of this model is that the decay patterns of the VLQ can be substantially altered from the usual scenario.  That is, the VLQ is a ``maverick top partner.''  As shown in Figs.~\ref{fig:TBRSorAD}(a,b), if the scale of the $U(1)_d$ is smaller than the EW sector ($v_d\lesssim v_{EW}$), the VLQ decays into a dark photon or dark Higgs greater than $95\%$ of the time independent of the VLQ mass.  This is due to the longitudinal enhancement of decaying into light gauge bosons which enhances the VLQ partial widths into $\gamma_d/h_2$ by $(v_{EW}/v_d)^2$ relative to decays into the SM EW bosons.  When the VLQ is substantially heavier than the top quark $M_T\gg M_t$, there can also an enhancement of $(M_T/M_t)^2$ for VLQ decays into $\gamma_d/h_2$~\cite{Rizzo:2018vlb}.

The appeal of this scenario is that the production rate of the dark photon $\gamma_d$ is largely independent of model parameters.  The VLQs can be pair produced via the strong interaction.  This pair production rate is governed by gauge interactions and only depends on the VLQ mass and spin.  As discussed above, the branching ratio ${\rm BR}(T\rightarrow t\,\gamma_d)=50\%$ in a very wide range of parameter space.  Hence, the dark photon production rate is almost completely governed the strong interaction and is independent of the small kinetic mixing parameter $\varepsilon$.

While the production rate of the dark photon is independent of the kinetic mixing parameter, the collider searches are not.  As we showed, for reasonable $\varepsilon$, the dark photon can give rise to displaced vertices, decay inside the detector, or even escape the detector and appear as missing energy as shown in Fig.~\ref{fig:Final}.  Besides the missing energy, the most promising signatures of the dark photon would be its decays into electrons and muons.  For dark photon masses much below the VLQ masses, the electrons and muons would be highly collimated giving rise to lepton jets~\cite{ArkaniHamed:2008qp} or even displaced lepton jets.

The model presented here is a mild perturbation from the typical simplified models of dark photons and VLQs.  However, as we demonstrated, the collider phenomenology is significantly changed from the usual scenarios.  Hence, this provides a robust framework in which searches for heavy particles at the LHC can illuminate a light dark sector force.

\section*{Acknowledgements}
We would like to thank KS Babu, KC Kong, and Chris Rogan for helpful discussions.  We would also like to thank KC Kong and Douglas McKay for suggestions the term ``maverick top partner.'' JHK and IML are supported in part by the U.S. Department of Energy under grant No.~DE-SC0017988.  MS is supported in part by the State of Kansas EPSCoR grant program.  SDL acknowledges financial support of Madison \& Lila Self Graduate Fellowship.  HSL is supported by National Research
Foundation (NRF) Strategic Research Program (NRF2017R1E1A1A01072736).  The data to reproduce the plots has been uploaded with the arXiv
submission or is available upon request.

\appendix
%%%%%%%%%%%%%%%%%%%%%%%%%%%%%%%%%%%%%%%%%%%%%%%%%%%%%%
\section{Perturbative Unitarity}
\label{app:PertUnit}
To derive a perturbative unitarity bound, we will look at tree-level $h_d t\rightarrow h_d t$ scattering in the high energy limit. For the high energy limit, we work with gauge eigenstate fields in the unbroken phase. In the broken phase, there would be $t$-channel diagrams with scalar trilinears.  However, since trilinear scales are dimensionful, they will be suppressed by $E^{-1}$ compared to the fermionic $s$-channel, were $E$ is the energy of the process.
\begin{figure}[tb]
\begin{center}
\includegraphics{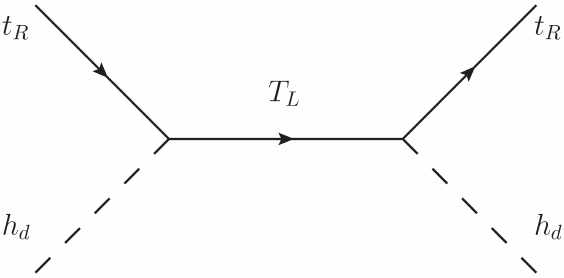} 
\caption{\label{fig:unitarityprocess} Tree-level diagram for perturbative unitarity bound of $\lambda_t$ from $h_d t\rightarrow h_d t$}
\end{center}
\end{figure}
Additionally, each vertex flips the chirality of the fermion. In the high energy limit, we neglect masses, so there can be no additional chiral flips from mass insertions.  Hence, the chirality of the incoming and outgoing top quarks must be the same.  Finally, the dark higgs vertex only exists between $t_R$ and $T_L$, so only the $t_R$ amplitudes are non-zero.   Fig.~\ref{fig:unitarityprocess} shows the relevant diagram.

The tree-level amplitude for this process is given by

\beq
i \mathcal{M}(h_d t_R \rightarrow h_d t_R) = -i \frac{\lambda_t^2}{2} \cos \frac{\theta}{2}~,
\eeq
where $\theta$ is the scattering angle. We now expand this amplitude into partial waves as

\beq
\mathcal{M} = 16 \pi \sum_{j=1/2,3/2,...} (2j+1) a_j d^j_{1/2,1/2}(\theta)~,
\eeq
where $d^j_{m,m'}(\theta)$ are Wigner d-functions. The only relevant term for this particular process corresponds to the $d^{1/2}_{1/2,1/2}(\theta)=\cos{\frac{\theta}{2}}$ term, and we get
\beq
a_{1/2}=-\frac{\lambda_t^2}{64 \pi}~.
\eeq
Tree-level perturbative unitarity then corresponds to $|{\rm Re}\,a_{1/2}| \leq \frac{1}{2}$, which, after taking a square root, gives us the final form of our bound,
\beq
|\lambda_t| \leq 4 \sqrt{2 \pi}~.
\eeq

%%%%%%%%%%%%%%%%%%%%%%%%%%%%%%%%%%%%%%%%%%%%%%%%%%%%%%
%%%%%%%%%%%%%%%%%%%%%%%%%%%%%%%%%%%%%%%%%%%%%%%%%%%%%%
\section{Kinetic Mixing}
\label{app:KineticM}
%%%%%%%%%%%%%%%%%%%%%%%%%%%%%%%%%%%%%%%%%%%%%%%%%%%%%%
%
We now review diagonalizing the neutral gauge bosons.  First, the kinetic mixing term in Eq.~(\ref{eq:GaugeK}) with the transformations
\begin{eqnarray}
B'_{\mu}= B_{\mu}+\frac{\varepsilon'}{\hat{c}_W\sqrt{1-{\varepsilon'}^2/\hat{c}_W^2}}B_{d, \mu},\quad  B'_{d, \mu}=\frac{1}{\sqrt{1-{\varepsilon'}^2/\hat{c}_W^2}}B_{d,\mu}~,
\end{eqnarray}
where $\hat{c}_W=\cos\hat{\theta}_W$. Then we have the normalized gauge kinetic terms
\begin{eqnarray}
\mathcal{L}_{\text{Gauge}}&\supset&-\frac{1}{4}B_{\mu\nu}{B}^{\mu\nu}-\frac{1}{4}B_{d,\mu\nu}B_d^{\mu\nu}~.
\end{eqnarray}
The covariant derivative in Eq.(\ref{eq:Cov}) then contains
\begin{eqnarray}
D_\mu&\supset&-i\,g\,(T^+W^++T^-W^-)-ig T^3W^3_\mu-ig' Y B_\mu-i (g_d Y_d+\frac{g'}{\hat{c}_W}\varepsilon Y) B_{d,\mu}~,
\end{eqnarray}
where we have defined $g_d=g'_d/\sqrt{1-{\varepsilon'}^2/\hat{c}_W^2}$, $\varepsilon=\varepsilon'/\sqrt{1-{\varepsilon'}^2/\hat{c}_W^2}$.% and
\begin{eqnarray}
T^+=\frac{1}{\sqrt{2}}\begin{pmatrix}0&1\\0&0 \end{pmatrix},\quad T^-=\frac{1}{\sqrt{2}}\begin{pmatrix}0&0\\1&0\end{pmatrix}~.
\end{eqnarray}
Here, $B$ is identified as the SM-like hypercharge gauge boson.  To get the photon we perform the usual rotation:
\begin{eqnarray}
\begin{pmatrix}W^3_{\mu}\\ B_{\mu} \end{pmatrix}=\begin{pmatrix}\hat{c}_W & \hat{s}_W \\ -\hat{s}_W & \hat{c}_W \end{pmatrix}\begin{pmatrix}\hat{Z}_{\mu} \\ A_{\mu} \end{pmatrix}~,
\end{eqnarray}
where $\hat{s}_W=\sin\hat{\theta}_W$.
The covariant derivative becomes
\begin{eqnarray}
D_\mu \supset -i\,g\,(T^+W^++T^-W^-)-i\,e\,Q\,A_\mu-i\,\hat{g}_Z\hat{Q}_Z\hat{Z}_\mu - i(g_d Y_d+\varepsilon \frac{g'}{\hat{c}_W} Y)B_{d,\mu}~, \nonumber
\end{eqnarray}
where $e=g\sin\hat\theta_W=g'\cos\hat\theta_W$,  $\hat{Q}_Z=T_3-\hat{x}_W Q$, $\hat{x}_W=\sin^2\hat{\theta}_W$, and $\hat{g}_Z=e/\cos\hat\theta_W/\sin\hat\theta_W$.  The charge operator is $Q=T_3+Y$.  Hence, $Q\,S=0$ and $Q\,\Phi=0$ in the unitary gauge.  We can identify $A_{\mu}$ as the physical massless photon and $e$ as the electric charge.

Evaluate the scalar kinetic terms to get the gauge boson masses:
\begin{eqnarray}
\mathcal{L}_{S,kin}&=&|D_\mu S|^2+|D_\mu \Phi|^2\nonumber\\
&\supset& M_W^2\,W^+_\mu W^{-,\mu}+\frac{1}{2}\left[\left(M_{\gamma_d}^0\right)^2B_{d,\mu}B_d^{\mu}-2\,M_{\gamma_dZ}^2\,\hat{Z}^\mu B_{d,\mu}+\left(M_Z^0\right)^2\hat{Z}^\mu\hat{Z}_\mu\right]~,
\end{eqnarray}
where
\begin{gather}
M_W=\frac{1}{2}gv =\frac{ev}{2 \sin \hat{\theta}_W },\quad\left(M_{\gamma_d}^0\right)^2=g_d^2\,v_d^2+\varepsilon^2\hat{t}_W^2\left(M_Z^0\right)^2,\nonumber\\
 M^2_{\gamma_d Z}=\varepsilon\,\hat{t}_W\,\left(M_Z^0\right)^2,\quad M_Z^0=\frac{1}{2}\hat{g}_Zv~,
\label{eq:GaugeMasses}
\end{gather}
and $\hat{t}_W=\hat{s}_W/\hat{c}_W$. We rotate the basis to diagonalize the mass matrix
\begin{eqnarray}
\begin{pmatrix}\hat{Z}^\mu \\B_d^\mu\end{pmatrix}=\begin{pmatrix} \cos\theta_d & \sin\theta_d\\-\sin\theta_d & \cos\theta_d\end{pmatrix}\begin{pmatrix}Z^\mu \\ \gamma^\mu_d\end{pmatrix}~,
\end{eqnarray}
where $Z$ has mass $M_Z$ and $\gamma_d$ has mass $M_{\gamma_d}$.  We can solve for the mixing angle and masses
\begin{eqnarray}
\sin(2\theta_d)&=&\frac{2\varepsilon\,\hat{t}_W\,(M_Z^0)^2}{M_Z^2-M_{\gamma_d}^2}~,\label{eq:thetaD}\\
M_Z^0&=&\sqrt{M_Z^2\cos^2\theta_d+M_{\gamma_d}^2\sin^2\theta_d}~,\label{eq:MZ0}\\
M_{\gamma_d}^0&=&\sqrt{M_{\gamma_d}^2\cos^2\theta_d+M_{Z}^2\sin^2\theta_d}~.
\end{eqnarray}

Now, for the degrees of freedom we choose:
\begin{eqnarray}
G_F,\quad \alpha_{EM}(0),\quad M_Z,\quad M_{\gamma_d},\quad \varepsilon,\quad v_d~,\label{eq:Parms}
\end{eqnarray}
with the $G_F,\alpha_{EM}(0),M_Z$ values in Eq.~(\ref{eq:EWParms}).  Using $v_{EW}^2=1/\sqrt{2}/G_F$ together with $g=e/\hat{s}_W$ and Eqs.~(\ref{eq:GaugeMasses},\ref{eq:MZ0}), we find
\begin{eqnarray}
\hat{c}_W^2=\frac{1}{2}+\frac{1}{2}\sqrt{1-\frac{2\sqrt{2}\alpha_{EM}(0)}{G_FM_Z^2}\frac{1}{\cos^2\theta_d+\tau_{\gamma_d}^2\sin^2\theta_d}}~\label{eq:cwhat},
\end{eqnarray}
where $\tau_{\gamma_d}=M_{\gamma_d}/M_Z$.  Now, Eqs.~(\ref{eq:thetaD},\ref{eq:cwhat}) can be used to recursively solve for $\sin\theta_d$ as an expansion in $\varepsilon$.  All other parameters can then be easily solved for in terms of the input parameters in Eq.~(\ref{eq:Parms}).

%%%%%%%%%%%%%%%%%%%%%%%%%%%%%%%%%%%%%%%%%%%%%%%%%%%%%%
\section{Scalar Interactions}
\label{app:Scalar}
%%%%%%%%%%%%%%%%%%%%%%%%%%%%%%%%%%%%%%%%%%%%%%%%%%%%%%
%
The scalar couplings relevant the analysis here are 
\begin{eqnarray} \label{eq:Scalar_Gauge}
\mathcal{L}_{V-S} &=& \frac{1}{1+\delta_{VV'}}\lambda_{h_i V V'}  h_i V_{\mu}{V'}^{\mu}  ~.
\end{eqnarray}
The detailed expressions for the couplings are 
\bea
\lambda_{h_1 W^+ W^-} &=& \frac{2 M^2_W}{v} \cos \theta_S~,\\
 \lambda_{h_2 W^+ W^-} &=& \frac{2 M^2_W}{v} \sin \theta_S~, \\
\lambda_{h_1ZZ}&=&\frac{2\,M_Z^2}{v}\cos\theta_S-2\,g_d^2\sin^2\theta_d\,v_d\left(\sin\theta_S+\cos\theta_S\,\tan\beta\right)~,\\
\lambda_{h_2ZZ}&=&\frac{2\,M_Z^2}{v}\sin\theta_S+2\,g_d^2\sin^2\theta_d\,v_d\left(\cos\theta_S-\sin\theta_S\,\tan\beta\right)~,\\
\lambda_{h_1\gamma_d\gamma_d}&=&\frac{2\,M_{\gamma_d}^2}{v}\cos\theta_S-2\,g_d^2\cos^2\theta_d\,v_d\left(\sin\theta_S+\cos\theta_S\,\tan\beta\right)~,\\
\lambda_{h_2\gamma_d\gamma_d}&=&\frac{2\,M_{\gamma_d}^2}{v}\sin\theta_S+2\,g_d^2\cos^2\theta_d\,v_d\left(\cos\theta_S-\sin\theta_S\,\tan\beta\right)~,\\
\lambda_{h_1\gamma_dZ}&=&g_d^2\,\sin2\theta_d\,v_d\left(\sin\theta_S+\cos\theta_S\,\tan\beta\right)~,\\
\lambda_{h_2\gamma_dZ}&=&-g_d^2\,\sin2\theta_d\,v_d\left(\cos\theta_S-\sin\theta_S\,\tan\beta\right)~,
\eea
where $M_W,M_Z^0$ are given in Eq.(\ref{eq:GaugeMasses}). The detailed expressions for gauge-sector parameters can be found in Appendix \ref{app:KineticM}. 

The $h_{1,2}$ trilinear are given by
\begin{eqnarray} \label{eq:Scalar_Self}
\mathcal{L}_{S-S-S} &=& \frac{1}{3!}\lambda_{h_1 h_1 h_1}  h^3_1   +\frac{1}{2}  \lambda_{h_1 h_1 h_2}  h^2_1 h_2  +  \frac{1}{2}\lambda_{h_1 h_2 h_2}  h_1 h^2_2 + \frac{1}{3!}\lambda_{h_2 h_2 h_2}  h^3_2~,
\end{eqnarray}
where the couplings are 
\begin{eqnarray}
\lambda_{h_1h_1h_1}&=&3\frac{M_1^2}{v}\left(\cos^3\theta_S-\frac{\sin^3\theta_S}{\tan\beta}\right)~,\\
\lambda_{h_1h_1h_2}&=&\frac{2\,M_1^2+M_2^2}{2\,v}\sin2\theta_S\left(\cos\theta_S+\frac{\sin\theta_S}{\tan\beta}\right)~,\\
\lambda_{h_1h_2h_2}&=&-\frac{M_1^2+2\,M_2^2}{2\,v}\sin2\theta_S\left(\frac{\cos\theta_S}{\tan\beta}-\sin\theta_S\right)~,\\
\lambda_{h_2h_2h_2}&=&3\frac{M_2^2}{v}\left(\frac{\cos^3\theta_S}{\tan\beta}+\sin^3\theta_S\right)~.
\end{eqnarray}

%%%%%%%%%%%%%%%%%%%%%%%%%%%%%%%%%%%%%%%%%%%%%%%%%%%%%%
\section{Gauge Boson Fermion Interactions}
\label{app:Dark}
%%%%%%%%%%%%%%%%%%%%%%%%%%%%%%%%%%%%%%%%%%%%%%%%%%%%%%
The $W$ interactions with the top quark and VLQ are
\begin{eqnarray}
\mathcal{L}_{W-f}=\frac{g}{\sqrt{2}}\cos\theta_L^t\,W^+_\mu\overline{t}\gamma^\mu\,P_L\,b+\frac{g}{\sqrt{2}}\sin\theta_L^t\,W^+_\mu\overline{T}\gamma^\mu P_Lb+{\rm h.c.}~.\label{eq:charged}
\end{eqnarray}
The couplings of the neutral bosons to top quark and VLQ are defined as
\begin{eqnarray}
\mathcal{L}_{V-f}=V_\mu\overline{f}\gamma^\mu\left(c^V_{L,ff'}P_L+c^V_{R,ff'}P_R\right)f'~,
\end{eqnarray}
where $V=\gamma_d,\,Z$ and the couplings are
\begin{eqnarray}
c^Z_{L,tt}&=&\hat{g}_Z\left[\cos\theta_d\left(\frac{1}{2}\cos^2\theta_L^t-Q_t\hat{s}^2_W\right)-\frac{1}{6}\sin\theta_d\hat{t}_W\,\varepsilon\left(1+3\,\sin^2\theta_L^t\right)\right]-g_d\,\sin\theta_d\,\sin^2\theta_L^t~,\nonumber\\
c^Z_{R,tt}&=&-Q_t\,\hat{g}_Z\left(\cos\theta_d\,\hat{s}_W^2+\varepsilon\,\sin\theta_d\,\hat{t}_W\right)-g_d\,\sin\theta_d\,\sin^2\theta_R^t~,\\
c^Z_{L,TT}&=&\hat{g}_Z\left[\cos\theta_d\left(-Q_t\,\hat{s}_W^2+\frac{1}{2}\sin^2\theta_L^t\right)-\sin\theta_d\,\hat{t}_W\varepsilon\left(Q_T-\frac{1}{2}\sin^2\theta_L^t\right)\right]-g_d\,\sin\theta_d\,\cos^2\theta_L^t~,\nonumber\\
c^Z_{R,TT}&=&-Q_T\,\hat{g}_Z\left(\cos\theta_d\,\hat{s}_W^2+\varepsilon\,\sin\theta_d\,\hat{t}_W\right)-g_d\,\sin\theta_d\,\cos^2\theta_R^t~,\\
c^Z_{L,Tt}&=&c^Z_{L,tT}=\frac{1}{4}\sin2\theta_L^t\left[\hat{g}_Z\left(\cos\theta_d+\sin\theta_d\,\hat{t}_W\varepsilon\right)+2\,g_d\,\sin\theta_d\right]~,\nonumber\\
c^Z_{R,Tt}&=&c^Z_{R,tT}=\frac{1}{2}g_d\,\sin\theta_d\,\sin2\theta_R^t~,\\
c^{\gamma_d}_{L,tt}&=&\hat{g}_Z\left[\sin\theta_d\left(\frac{1}{2}\cos^2\theta_L^t-Q_t\hat{s}_W^2\right)+\frac{1}{6}\cos\theta_d\,\hat{t}_W\,\varepsilon\left(1+3\,\sin^2\theta_L^t\right)\right]+g_d\,\cos\theta_d\,\sin^2\theta_L^t~,\nonumber\\
c^{\gamma_d}_{R,tt}&=&-Q_t\,\hat{g}_Z\left(\sin\theta_d\,\hat{s}^2_W-\cos\theta_d\,\hat{t}_W\,\varepsilon\right)+g_d\,\cos\theta_d\,\sin^2\theta_R^t~,\\
c^{\gamma_d}_{L,TT}&=&\hat{g}_Z\left[\sin\theta_d\left(-Q_t\,\hat{s}_W^2+\frac{1}{2}\sin^2\theta_L^t\right)+Q_t\,\cos\theta_d\,\hat{t}_W\varepsilon\left(1-\frac{3}{4}\sin^2\theta_L^t\right)\right]+g_d\,\cos\theta_d\,\cos^2\theta_L^t~,\nonumber\\
c^{\gamma_d}_{R,TT}&=&-Q_t\,\hat{g}_Z\left(\sin\theta_d\,\hat{s}^2_W-\cos\theta_d\,\hat{t}_W\varepsilon\right)+g_d\,\cos\theta_d\,\cos^2\theta_R^t~,\\
c^{\gamma_d}_{L,Tt}&=&c^{\gamma_d}_{L,tT}=\frac{1}{4}\sin2\theta_L^t\left[\hat{g}_Z\left(\sin\theta_d-\cos\theta_d\hat{t}_W\varepsilon\right)-2\,g_d\,\cos\theta_d\right]~,\nonumber\\
c^{\gamma_d}_{R,Tt}&=&c^{\gamma_d}_{R,tT}=-\frac{1}{2}g_d\,\sin2\theta_R^t\cos\theta_d~.
\end{eqnarray}
The $\gamma_d,\,Z$ interactions with the other SM fermions are flavor diagonal and can be obtained using the covariant derivative defined in Eq.(\ref{eq:NewCovar}) and their SM quantum numbers.

%%%%%%%%%%%%%%%%%
% References
%%%%%%%%%%%%%%%%%
% change_____________________
%\bibliographystyle{plain}
\bibliographystyle{myutphys}
% _____________________
\bibliography{bibliography}

\end{document}